\pdfoutput=1  
\documentclass[acmsmall]{acmart}

\usepackage{tabularx}
\usepackage{blindtext}
\usepackage{amsmath}
\usepackage{amsfonts}

\usepackage{pgfplots}   
\usepackage{tikz}       
\usepackage{subcaption} 
\usepackage{graphicx}
\usepackage{textcomp}
\usepackage{color}
\usepackage{xcolor}
\usepackage{multirow}
\usepackage{enumitem}
\usepackage{dsfont}
\usepackage[linesnumbered,ruled,vlined]{algorithm2e}
\usepackage{balance}
\AtBeginDocument{%
  }

\setcopyright{acmlicensed}
\acmJournal{PACMMOD}
\acmYear{2025} \acmVolume{0} \acmNumber{0} \acmArticle{xxx} \acmMonth{0}\acmDOI{xxx}

\begin{document}
\newcommand{\kw}[1]{{\ensuremath {\mathsf{#1}}}\xspace}
\newcommand{\stitle}[1]{\vspace{1ex} \noindent{\textbf{#1}}}

\SetKwFunction{FTriangles}{UpdateTriangles}
\SetKwFunction{FTrianglesTwo}{UpdateTriangles2}
\SetKwFunction{Pedges}{ProcessHyperedge}
\SetKwFunction{Sedges}{SampleHyperedge}
\SetKwProg{Fn}{Function}{:}{}

\definecolor{c1}{RGB}{42,99,172} %
\definecolor{c2}{RGB}{255,88,93}
\definecolor{c3}{RGB}{255,181,73}
\definecolor{c4}{RGB}{119,71,64} %
\definecolor{c5}{RGB}{228,123,121} %
\definecolor{c6}{RGB}{208,167,39} %
\definecolor{c7}{RGB}{0,51,153}
\definecolor{c8}{RGB}{56,140,139} 
\definecolor{c9}{RGB}{0,0,0} 

\newcommand{\reffig}[1]{Figure~\ref{fig:#1}}
\newcommand{\refsec}[1]{Section~\ref{sec:#1}}
\newcommand{\reftable}[1]{Table~\ref{tab:#1}}
\newcommand{\refalg}[1]{Algorithm~\ref{alg:#1}}
\newcommand{\refeq}[1]{Equation~\ref{eq:#1}}
\newcommand{\refdef}[1]{Definition~\ref{def:#1}}

\newcommand{\twitterut}{\textit{Twitter-ut}\xspace}
\newcommand{\editfrwiki}{\textit{Edit-frwiki}\xspace}
\newcommand{\amazonratings}{\textit{AmazonRatings}\xspace}
\newcommand{\movielens}{\textit{Movie-lens}\xspace}
\newcommand{\edititwiki}{\textit{Edit-itwiki}\xspace}
\newcommand{\lastfmband}{\textit{Lastfm-band}\xspace}
\newcommand{\discogs}{\textit{Discogs}\xspace}
\newcommand{\yahoosongs}{\textit{Yahoo-songs}\xspace}
\newcommand{\stackoverflow}{\textit{StackoverFlow}\xspace}
\newcommand{\livejournal}{\textit{LiveJournal}\xspace}
\newcommand{\deliciousui}{\textit{Delicious-ui}\xspace}
\newcommand{\orkut}{\textit{Orkut}\xspace}

\newcommand{\deabc}{\kw{DEABC}}
\newcommand{\deabcpro}{\kw{DEABC^+}}
\newcommand{\abacus}{\kw{ABACUS}}
\newcommand{\fleet}{\kw{FLEET3}}
\newcommand{\cas}{\kw{CAS{-}R}}

\newcommand{\ma}{\textit{MAG}\xspace}
\newcommand{\walmart}{\textit{Walmart}\xspace}
\newcommand{\ndc}{\textit{NDC}\xspace}
\newcommand{\tc}{\textit{Trivago-clicks}\xspace}
\newcommand{\cb}{\textit{Congress-bills}\xspace}
\newcommand{\mg}{\textit{MAG-Geology}\xspace}
\newcommand{\dblp}{\textit{DBLP}\xspace}
\newcommand{\ts}{\textit{Threads-stack}\xspace}

\newcommand{\al}{\kw{HTCount}}
\newcommand{\alp}{\kw{HTCount}-\kw{P}}
\newcommand{\hypersv}{\kw{HyperSV}}
\newcommand{\hyperwsv}{\kw{HyperWSV}}
\newcommand{\hypershe}{\kw{HyperSHE}}
\newcommand{\hypersvars}{\kw{HyperSV}-\kw{ARS}}

\newcommand{\todo}[1]{\textcolor{red}{$\Rightarrow:$ #1}}
\newcommand{\revise}[1]{{#1}}

\newcommand*\circled[1]{\tikz[baseline=(char.base)]{
            \node[shape=circle,draw,inner sep=1pt] (char) {#1};}}

\newcommand{\reviseone}[1]{{#1}}
\newcommand{\revisetwo}[1]{{#1}}   
\newcommand{\revisethree}[1]{{#1}} 
\title{Triangle Counting in Hypergraph Streams: A Complete and Practical Approach}


\author{Lingkai Meng}
\affiliation{%
  \institution{Shanghai Jiao Tong University}
  \city{Shanghai}
  \country{China}}
\email{mlk123@sjtu.edu.cn}

\author{Long Yuan}
\authornote{Corresponding author.}
\affiliation{%
  \institution{Wuhan University of Technology}
  \city{Wuhan}
  \country{China}}
\email{longyuanwhut@gmail.com}

\author{Xuemin Lin}
\affiliation{%
  \institution{Shanghai Jiao Tong University}
  \city{Shanghai}
  \country{China}}
\email{xuemin.lin@sjtu.edu.cn}

\author{Wenjie Zhang}
\affiliation{%
  \institution{University of New South Wales}
  \city{Sydney}
  \country{Australia}}
\email{wenjie.zhang@unsw.edu.au}

\author{Ying Zhang}
\affiliation{%
  \institution{Zhejiang Gongshang University}
  \city{Hangzhou}
  \country{China}}
\email{ying.zhang@zjgsu.edu.cn}

\renewcommand{\shortauthors}{Lingkai Meng et al.}

\begin{abstract}
Triangle counting in hypergraph streams—including both hyper-vertex and hyper-edge triangles—is a fundamental problem in hypergraph analytics, with broad applications. However, existing methods face two key limitations: $(i)$ an incomplete classification of hyper-vertex triangle structures, typically considering only inner or outer  triangles; and $(ii)$ inflexible sampling schemes that predefine the number of sampled hyperedges, which is impractical under strict memory constraints due to highly variable hyperedge sizes.
To address these challenges, we first introduce a complete classification of hyper-vertex triangles, including inner, hybrid, and outer triangles. Based on this, we develop \al, a reservoir-based algorithm that dynamically adjusts the sample size based on the available memory $M$. To further improve memory utilization and reduce estimation error, we develop \alp, a partition-based variant that adaptively partitions unused memory into independent sample subsets.
We provide theoretical analysis of the unbiasedness and variance bounds of the proposed algorithms. Case studies demonstrate the expressiveness of our triangle structures in revealing meaningful interaction patterns. Extensive experiments on real-world hypergraphs show that both our algorithms achieve highly accurate triangle count estimates under strict memory constraints, with relative errors that are 1 to 2 orders of magnitude lower than those of existing methods and consistently high throughput.
\end{abstract}


\received{April 2025}
\received[revised]{July 2025}
\received[accepted]{August 2025}

\maketitle

\section{Introduction}
\label{sec:intro}

\begin{figure}[t]
    \centering
        \centering
            \begin{subfigure}{0.35\textwidth}
        \centering
        \includegraphics[width=0.55\linewidth]{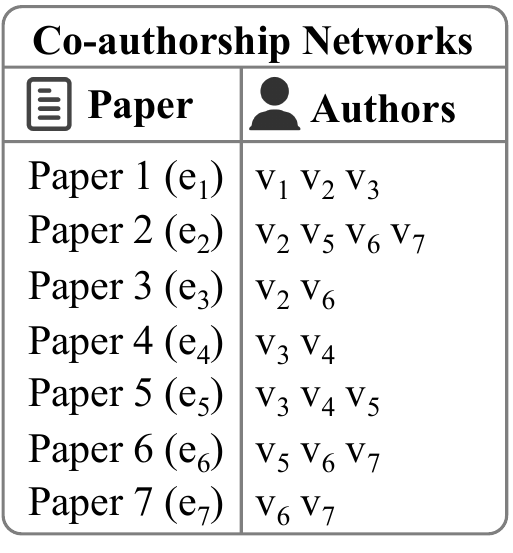}
        \caption{A Co-authorship Network}
    \end{subfigure}
    \hspace{0.3cm}
    \begin{subfigure}{0.35\textwidth}
        \centering
        \includegraphics[width=0.62\linewidth]{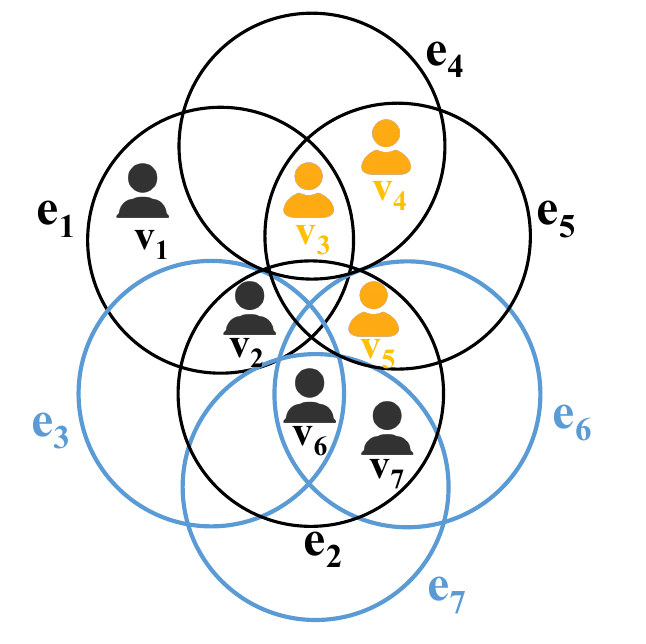}
        \caption{The Corresponding Hypergraph}
    \end{subfigure}
    \vspace{-0.2cm}
        \caption{A Hypergraph Example}
        \label{fig:hypergraph}
         \vspace{-0.5cm}
\end{figure}

A hypergraph is a generalization of a traditional graph~\cite{DBLP:journals/dase/ChenFYLW23,DBLP:conf/www/TurkT19,
      DBLP:conf/dac/WangYZQLCJCQZ20,luo2025efficient,DBLP:journals/vldb/LeeSF20,meng2022index,xu2007scan,chen2019efficient,gao2023efficient,gao2017efficiently,zhang2024efficient,zeng2021efficient} that allows a hyperedge to connect any number of vertices, which has attracted extensive research attention~\cite{chen2018scalable,yin2025efficient,zhang2023efficiently,lotito2022higher,luo2024hierarchical}. This structure naturally captures the many-to-many interactions found in a wide range of real-world systems, such as social networks~\cite{zhang2025accelerating,li2013link,zhu2018social,liu2020vac}, biological networks~\cite{yang2023hgmatch,lugo2021classification,feng2021hypergraph}, collaborative shopping networks~\cite{han2023search,wu2022hypergraph}, and co-authorship networks~\cite{yin2025efficient,inoue2022hypergraph}. 
Figure~\ref{fig:hypergraph}(a) shows a co-authorship network as a table, with its corresponding hypergraph in Figure~\ref{fig:hypergraph}(b). Each hyperedge (circle) represents a paper, and its vertices correspond to the authors. Compared with general graphs, we can intuitively observe collaboration relationships in hypergraphs through shared authors.
In many practical scenarios, hypergraphs are dynamic, evolving dynamically as hypergraph streams, where hyperedges arrive continuously at high velocity and potentially unbounded volume \cite{zhang2023efficiently,alistarh2015streaming,reinstadtler2025semi}. This streaming nature makes traditional offline analysis methods impractical due to prohibitive memory requirements and computational delay. 
Consequently, streaming hypergraph analytics, designed to efficiently estimate key structural patterns and statistics under strict memory and latency constraints, has emerged as a critical research area \cite{reinstadtler2025semi,tacsyaran2021streaming,kurte2021phoenix}.

Among various streaming hypergraph analytics tasks, triangle counting, a cornerstone of traditional graph analysis \cite{meng2025revisiting,meng2024survey,DBLP:conf/www/TurkT19,
      DBLP:conf/dac/WangYZQLCJCQZ20,
      DBLP:journals/vldb/LeeSF20,
      DBLP:journals/tpds/PandeyWZTZLLHDL21,
      DBLP:journals/tpds/YasarRBC22,
      DBLP:conf/kdd/AhmedDNK14,
      DBLP:journals/ac/RavichandranSAK23,DBLP:conf/kdd/TsourakakisKMF09}, 
holds particular importance in the context of hypergraphs. Efficiently identifying and counting these triangles in hypergraph streams is essential for uncovering latent community structures, capturing higher-order interactions, and gaining insights into the organization of dynamic, complex systems \cite{10.14778/3407790.3407823,zhang2023efficiently,wang2022efficient}. Such analyses have demonstrated applications across diverse domains.  For example: $(i)$ \emph{Network Analysis.} In academic collaboration networks, distinct triangle structures can highlight different teamwork patterns, and analyzing the distribution and frequency of these patterns can help uncover core research teams and influential individuals~\cite{yin2025efficient,wu2022hypergraph,juul2024hypergraph}.
$(ii)$ \emph{Clustering Coefficients.} Triangle counting can serve as a critical step in calculating clustering coefficients, which is defined as $3 \times \frac{|\triangle|}{|\sqcup|}$ where $|\triangle|$ is the number of hyper-edge triangles, and $|\sqcup|$ is the number of connected hyperedge pairs~\cite{purkait2016clustering,papa2007hypergraph}. 
$(iii)$ \emph{Trend Forecasting.} By tracking changes in triangle counts in streaming hypergraphs, one can effectively capture active periods and emerging research topics in scientific domains~\cite{lee2023temporal,lotito2022higher}.
\revisetwo{$(iv)$ \emph{Join Size Estimation in Databases.} Triangle counting in hypergraphs provides a principled way to estimate the size of multi-way joins in databases. By modeling tables and join queries as hypergraphs, the number of hyperedge triangles directly corresponds to the expected output size of a three-way join~\cite{atserias2013size,kallaugher2018sketching}, which supports query optimization and resource allocation in large-scale database systems.}

\stitle{Motivation.} Although triangle counting in hypergraph streams has attracted increasing attention~\cite{yin2025efficient,zhang2023efficiently,lotito2022higher}, the research remains fragmented and demonstrates deficiencies in both the modeling and algorithmic aspects: 

\begin{figure}[]
    \centering


    
    \begin{minipage}{0.35\textwidth}
        \centering
        \includegraphics[width=0.73\textwidth]{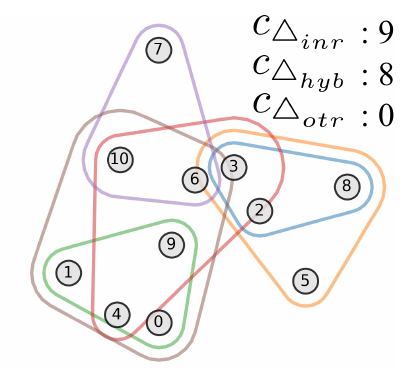}
        \subcaption{\mg Subgraph}
    \end{minipage}
    \hspace{0.05\textwidth}
    \begin{minipage}{0.35\textwidth}
        \centering
        \includegraphics[width=\textwidth]{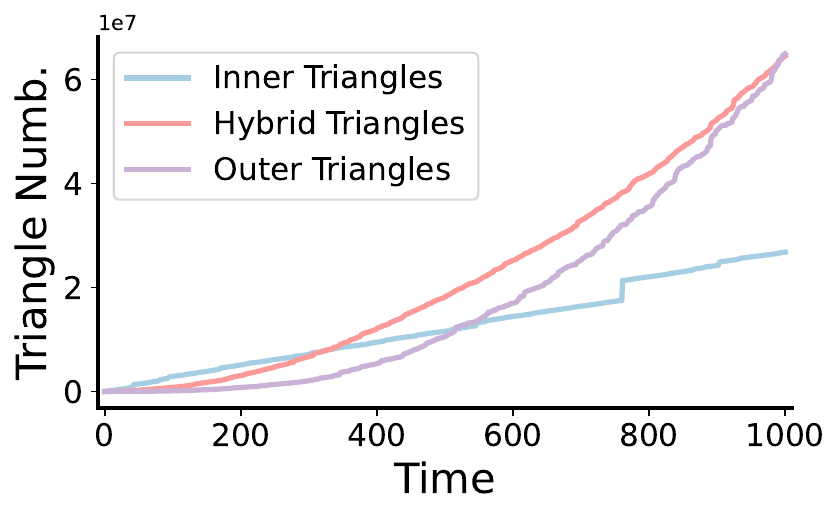}
        \subcaption{\mg Stream}
    \end{minipage}
    \caption{\reviseone{Case Studies of the Co-authorship Network}}
    \label{fig:case_intro}
\end{figure}

\begin{itemize}[leftmargin=*]
    \item \emph{\underline{Incomplete Model Taxonomy:}} According to the definition of hypergraphs, triangles in hypergraphs can be categorized into two types: hyper-vertex triangles (three vertices with mutual interactions like $\{v_3, v_4, v_5\}$ in Figure~\ref{fig:hypergraph}(b)) and hyper-edge triangles (three hyperedges that are pairwise connected through shared vertices like $\{e_3, e_6, e_7\}$ in Figure~\ref{fig:hypergraph}(b))\footnote{Other works may refer to these triangle structures using different terminologies. For example, hyper-vertex triangles are sometimes called “higher-order motifs”~\cite{lee2024survey} and hyper-edge triangles may be referred to as “hypergraph motifs”~\cite{lee2024survey} or “hyper-triangles”~\cite{yin2025efficient}. However, for consistency and clarity, we adopt the unified naming convention of “hyper-vertex triangles” and “hyper-edge triangles” in this paper.}. Existing literature provides a systematic classification of hyper-edge triangles into four distinct classes encompassing 20 patterns, followed by extensive quantitative investigations of these patterns \cite{yin2025efficient,10.14778/3407790.3407823}. In contrast, studies on hyper-vertex triangles remain conspicuously absent. Current classifications are rudimentary and fail to fully leverage the analytical potential of hyper-vertex triangles in hypergraph analysis.
    \reviseone{This analytical gap is vividly illustrated by the \mg co-authorship network. As shown in Figure~\ref{fig:case_intro}, both the static and dynamic perspectives reveal the unique and dominant role of hybrid triangles. In a representative subgraph (Figure~\ref{fig:case_intro}(a)), hybrid triangles appear almost as frequently as inner triangles, while outer triangles are entirely absent. This highlights the prevalence of overlapping collaborations between research teams—structures that cannot be captured by models considering only inner or outer triangles. 
    Besides, the evolution of triangle counts over time (Figure~\ref{fig:case_intro}(b)) demonstrates the dominance of hybrid triangles, which suggests cross-team and interdisciplinary collaborations are a fundamental and enduring feature in the development of the \mg community. More importantly, hybrid triangles serve as a sensitive indicator of structural change. Inner triangles dominate in the early stages, but a rapid increase in hybrid triangles marks the emergence of interdisciplinary collaboration. The subsequent rise of outer triangles reflects a shift toward broader cross-team interactions. Notably, the early surge in hybrid triangles provides a timely signal of research convergence and collaboration trends that would be overlooked by considering only inner or outer triangles. A more detailed analysis is provided in Section~\ref{sec:casestudy}.}
    
    
    \item \emph{\underline{Impractical Streaming Algorithm:}} In the literature, a sampling-and-estimation based method, named \hypersv, for counting triangles in hypergraphs has been proposed  \cite{zhang2023efficiently}. However, it is practically inapplicable  due to the following reasons: (1) \hypersv overlooks detailed pattern distinctions during the counting process and fails to provide counts for each specific pattern. This limitation restricts its utility in scenarios that require fine-grained analysis of triangle types. (2) \hypersv samples $\lambda$ edges to estimate the number of triangles in a hypergraph stream and assumes that the number of sampled edges $\lambda$ is given. While this assumption is reasonable for traditional graphs --- where the size of each edge is fixed and the number of sampled edges can be directly computed based on available memory $M$  --- it becomes problematic in hypergraphs. In hypergraphs, a hyperedge can connect any number of vertices, resulting in highly variable hyperedge sizes. As demonstrated in \reftable{datasets} in \refsec{ee}, the largest hyperedge size can be up to 200 times greater than the smallest hyperedge size. Consequently, determining an appropriate value of $\lambda$ is difficult for end users. Even if we ignore the practical constraint that the hyperedge information in the stream is unknown beforehand and assume that $|e|_{\min}$ and $|e|_{\max}$ are known, where $|e|_{\min}$ and $|e|_{\max}$ denote the smallest and largest hyperedge size, respectively. An optimistic strategy (i.e., $\lambda = \frac{M}{|e|_{\min}}$) risks exceeding the available memory $M$, while a pessimistic strategy (i.e., $\lambda = \frac{M}{|e|_{\max}}$) may result in substantial estimation errors due to underutilization of available memory.
    As verified in our experiments (\refsec{pe}), the optimistic strategy fails on all datasets due to memory overflow, while the pessimistic strategy leads to estimation errors that are 1–2 orders of magnitude higher.
    \revisethree{Moreover, while Al-Kateb et al.~\cite{al2007adaptive} have proposed an adaptive-size reservoir sampling method to dynamically adjust the sample size, directly applying their technique to hypergraph triangle counting introduces new challenges. Theoretically, when the sample size increases, their method cannot guarantee strictly unbiased estimation, leading to potential bias with highly variable data. From the implementation perspective, hyperedge sizes can vary greatly, making it difficult to predict the appropriate amount by which the sample size should be increased as the stream evolves.}
\end{itemize}

Motivated by these gaps, this paper conducts a comprehensive study on triangle counting in hypergraph streams with the objectives: (1) to investigate and classify hyper-vertex triangles, thereby establishing a complete taxonomy of triangle types in hypergraphs; and (2) to design a practical streaming algorithm for triangle counting that accurately estimates various patterns related to hyper-vertex triangles and hyper-edge triangles by efficiently utilizing available memory while maintaining low computational latency.

\stitle{Challenges.} Achieving these goals presents several key challenges: 

\begin{itemize}[leftmargin=*]
    \item For the classification of hyper-vertex triangles, the model must adhere to the fundamental principles of hypergraph theory while effectively capturing the diverse structural characteristics observed in real-world hypergraphs. 
    
    \item For the practical streaming algorithm, challenges arise from the constraints of the streaming model and the structural complexity of hypergraphs.  Limited available memory restricts the ability to preserve global topology, which is essential for accurate triangle estimation. Moreover, the high velocity of hyperedge arrivals requires real-time processing for each update. These challenges are further compounded by hypergraph-specific characteristics: the variable sizes of hyperedges render traditional sampling strategies ineffective. Adaptive mechanisms are therefore required to dynamically optimize memory utilization across hyperedges of varying sizes. Furthermore, the algorithm must identify and count over 20 distinct triangle patterns, each requiring specialized recognition and counting heuristics, which further increases the solution's overall complexity.
\end{itemize}

\stitle{Our solutions.} We address all of these challenges in this paper:

\begin{itemize}[leftmargin=*]

\item \emph{\underline{\revisethree{Comprehensive Taxonomy of Triangles in Hypergraphs:}}} \revisethree{Different from the existing studies that focus only on hyper-edge triangles~\cite{yin2025efficient} or provide incomplete classifications of hyper-vertex triangles~\cite{zhang2023efficiently}, typically omitting the important hybrid triangle structure, we propose the first complete and systematic taxonomy of hyper-vertex triangles in hypergraphs. Specifically, we categorize hyper-vertex triangles into three distinct types: inner triangles (three vertices are included in the same hyperedge), hybrid triangles (three vertices are contained in one hyperedge, while two of these vertices are also contained in another hyperedge), and outer triangles (three vertices are pairwise contained in three different hyperedges).
Our case studies in \refsec{casestudy} explicitly demonstrate that these newly defined hybrid triangles not only dominate among hyper-vertex triangles but also uncover meaningful and previously overlooked interaction patterns. 
To the best of our knowledge, this is the first work to establish a theoretically complete classification of hyper-vertex triangles, thereby bridging a key gap in hypergraph analysis.
}



\item \emph{\underline{\revisethree{Practical Streaming Algorithms:}}}
\revisethree{
We propose a unified computational framework for triangle counting in hypergraph streams. Unlike \hypersv that cannot determine the appropriate number of edges to sample under memory constraints, our method, \al, defines the sampling size based on the available memory $M$ directly and adaptively adjusts the number of sampled hyperedges.
Specifically, for each incoming hyperedge, we employ reservoir sampling; if the sample exceeds the memory budget, we iteratively evict hyperedges until sufficient space is available. After adding a new hyperedge, we identify all triangle types and update their counts using correction factors derived from the current sampling probability, ensuring unbiased estimation.
To further enhance memory utilization and estimation robustness, we introduce a partition-based algorithm (\alp), which dynamically splits unused memory into independent sample subsets. Each subset independently applies the same hyperedge sampling strategy, and incoming hyperedges are routed to subsets based on the weighted size of each subset. 
Unlike adaptive-size reservoir sampling~\cite{al2007adaptive} that simply increases sample size and may introduce bias under data variability, our method distributes surplus memory across subsets, strictly guaranteeing unbiasedness and efficient memory use.
We provide a theoretical analysis of the unbiasedness and variance bounds of our new algorithms, and experimental results demonstrate the superiority of our approach compared to SOTA algorithms.
}

\end{itemize}
\stitle{Contributions.} We make the following contributions in this paper:

\begin{itemize}[leftmargin=*]


\item[$\bullet$] \revisethree{We propose a comprehensive taxonomy for triangles in hypergraphs by completing the classification of hyper-vertex triangles. Together with the existing taxonomy of hyper-edge triangles, this forms a unified and complete framework for classifying all triangle types in hypergraphs.}

\item[$\bullet$] We propose a reservoir-based algorithm that dynamically adjusts the sampled set of hyperedges under a fixed available memory budget $M$. We then design a partition-based variant to further improve memory utilization and reduce variance. Both algorithms provide unbiased estimation of multiple types of triangles over hypergraph streams.

\item[$\bullet$] We provide a rigorous theoretical analysis for both algorithms, proving that the triangle count estimations are unbiased and have bounded variance. We also analyze their time and space complexity.

\item[$\bullet$] We conduct extensive experiments to evaluate the effectiveness and efficiency of our proposed algorithms. Our case studies demonstrate that the defined hyper-vertex triangle structures reveal meaningful interaction patterns in real hypergraphs. The performance results show that our algorithms achieve highly accurate triangle count estimates under strict memory constraints, achieving relative errors 1---2 orders of magnitude lower than existing methods and consistently high throughput.
\end{itemize}

\vspace{-0.2cm}
\section{Related Work}
\vspace{-0.1cm}
\label{sec:related_work}

\stitle{Triangle Counting over Static Graphs.}
A wide range of exact and approximate algorithms have been developed for triangle counting in traditional static graphs~\cite{DBLP:conf/kdd/TsourakakisKMF09,
      DBLP:conf/focs/EdenLRS15,
      DBLP:journals/tkde/WuYL16,
      DBLP:conf/soda/KallaugherP17,
      DBLP:journals/tpds/BissonF17,
      DBLP:conf/sc/HuLH18,
      DBLP:conf/www/TurkT19,
      DBLP:conf/dac/WangYZQLCJCQZ20,
      DBLP:journals/vldb/LeeSF20,
      DBLP:journals/tpds/PandeyWZTZLLHDL21,
      DBLP:journals/tpds/YasarRBC22,
      DBLP:conf/kdd/AhmedDNK14,
      DBLP:journals/tcs/Latapy08,
      DBLP:journals/ac/RavichandranSAK23,
      schank2005finding}.
Early traversal-based methods~\cite{DBLP:journals/tcs/Latapy08,schank2005finding} introduce optimized vertex iteration and ordering techniques for triangle counting. AOT~\cite{yu2020aot} refines traditional orientation frameworks by traversing based on vertex out-degrees, achieving performance and optimal theoretical complexity.
To address the high cost of exact counting, various sampling-based approximation methods have been proposed, including edge-based~\cite{DBLP:conf/kdd/TsourakakisKMF09,DBLP:conf/focs/EdenLRS15,DBLP:journals/tkde/WuYL16,DBLP:conf/kdd/AhmedDNK14}, wedge-based~\cite{DBLP:conf/www/TurkT19}, and hybrid~\cite{DBLP:conf/soda/KallaugherP17} approaches.
Recent advances also leverage hardware acceleration, such as GPU-based~\cite{DBLP:journals/tpds/PandeyWZTZLLHDL21,DBLP:journals/tpds/BissonF17,DBLP:conf/sc/HuLH18,DBLP:journals/tpds/YasarRBC22} and SIMD-based~\cite{DBLP:journals/ac/RavichandranSAK23} algorithms, as well as in-memory architectures like TCIM~\cite{DBLP:conf/dac/WangYZQLCJCQZ20}.
However, these methods are tailored to static, traditional graphs and do not extend to the richer structure of hypergraphs. For hypergraphs, Yin et al.~\cite{yin2025efficient} introduced a taxonomy of hyper-edge triangle patterns and a two-step framework based on hyperwedges for efficient and accurate triangle counting.

\stitle{Triangle Counting over Streaming Graphs.}
TRIÈST~\cite{stefani2017triest} introduces a family of reservoir-sampling algorithms for estimating both local and global triangle counts in fully dynamic streams. MASCOT~\cite{lim2015mascot} adopts a memory-aware sampling scheme to reduce estimation variance under constrained space, while Jha et al.~\cite{DBLP:conf/kdd/JhaSP13} propose a space-efficient technique inspired by the birthday paradox.
To address edge duplication and temporal constraints, sliding-window~\cite{DBLP:conf/sigmod/Gou021} and duplicate-aware~\cite{meng2024counting,DBLP:journals/pvldb/WangQSZTG17} algorithms have been proposed.
In addition, distributed, parallel, and hardware-accelerated methods~\cite{DBLP:journals/tkdd/YangSYGL22,DBLP:conf/srds/XuanLCLYF24,DBLP:journals/kbs/YangSGLL23,DBLP:journals/tpds/HuangWFWC22} have further improved scalability and throughput for large-scale streaming graphs.

All these methods target triangle counting in traditional graph streams and cannot be directly applied to hypergraph streams, where triangle structures are more complex and hyperedge sizes vary. The only existing work for hypergraph streams~\cite{zhang2023efficiently} has several limitations: it \emph{\underline{only considers inner and outer triangles}}, ignoring the important hybrid triangles; it does not distinguish patterns among hyper-edge triangles; and it assumes a fixed sample size, which cannot accommodate variable hyperedge sizes and may lead to memory inefficiency or estimation errors.

\stitle{\revisethree{Reservoir Sampling Techniques over Streaming Graphs.}}
\revisethree{Reservoir sampling is a classic technique for maintaining representative samples in streaming settings under memory constraints~\cite{vitter1985random}. It is widely used in streaming graph algorithms for tasks such as triangle~\cite{stefani2017triest,lim2015mascot} and butterfly counting~\cite{meng2024counting}, and has been extended to handle dynamic graphs with edge insertions and deletions~\cite{papadias2024counting,shin2020fast}.
To further improve memory utilization in streams with varying size, Al-Kateb et al.~\cite{al2007adaptive} introduced an adaptive-size reservoir sampling method that dynamically adjusts the reservoir size based on the observed stream.  It increases sample size by selectively incorporating new tuples and probabilistically retaining existing ones, ensuring that the overall uniformity confidence exceeds a user-defined threshold. However, traditional reservoir sampling and its variants typically require a predefined sample size, which is impractical for hypergraphs. Although adaptive-size reservoir sampling improves flexibility by allowing dynamic adjustment, it still cannot guarantee strict unbiasedness, potentially introducing estimation bias in the presence of significant data variability.
}

\vspace{-0.1cm}
\section{Problem Definition}
\label{sec:pd}

\begin{figure}[t]
    \centering
    \begin{subfigure}{0.26\textwidth}
        \centering
        \includegraphics[width=0.75\linewidth]{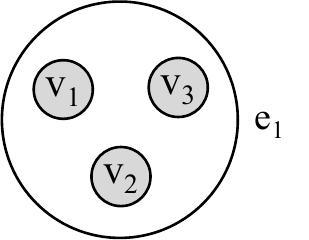}
        \caption{Inner Triangle}
    \end{subfigure}
    \begin{subfigure}{0.27\textwidth}
        \centering
        \includegraphics[width=1\linewidth]{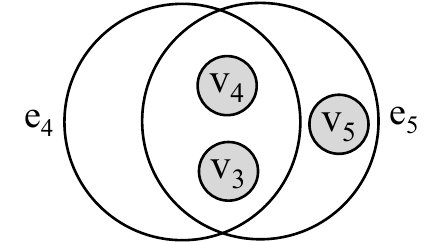}
        \caption{Hybrid Triangle}
    \end{subfigure}
    \hspace{0.1cm}
    \begin{subfigure}{0.26\textwidth}
        \centering
        \includegraphics[width=0.65\linewidth]{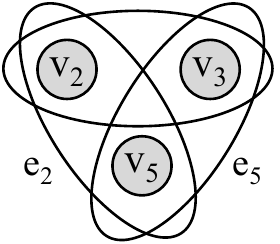}
        \caption{Outer Triangle}
    \end{subfigure}
    \caption{Hyper-vertex Triangles}
    \label{fig:Hyper-vertex Triangles}
\end{figure}

\begin{figure}[]
    \centering

    \begin{subfigure}{0.2\textwidth}
        \centering
        \includegraphics[width=0.9\linewidth]{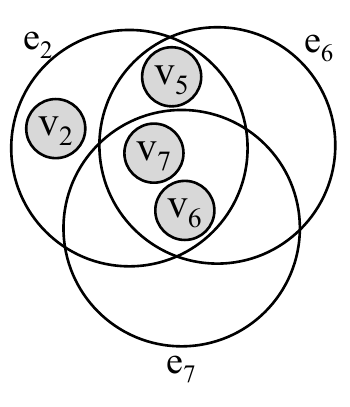}
        \caption{CCC}
    \end{subfigure}
    \begin{subfigure}{0.2\textwidth}
        \centering
        \includegraphics[width=0.9\linewidth]{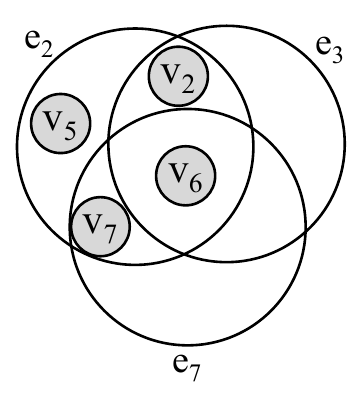}
        \caption{TCC}
    \end{subfigure}
    \begin{subfigure}{0.2\textwidth}
        \centering
        \includegraphics[width=0.9\linewidth]{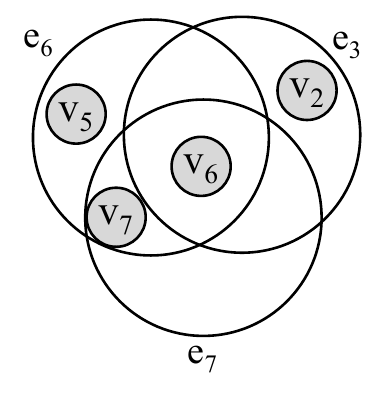}
        \caption{TTC}
    \end{subfigure}
    \begin{subfigure}{0.2\textwidth}
        \centering
        \includegraphics[width=0.9\linewidth]{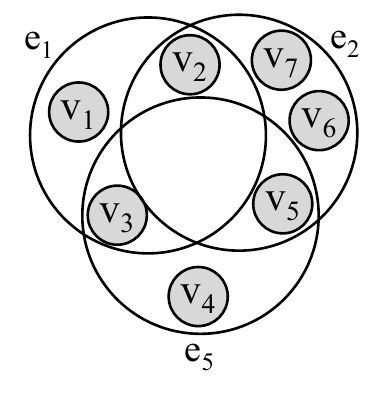}
        \caption{TTT}
    \end{subfigure}
    \caption{Hyper-edge Triangles}
    \label{fig:Hyper-edge Triangles}
\end{figure}

A hypergraph $H=(V,E)$ is defined as a graph where $V$ is the set of vertices and $E$ is the set of hyperedges, where each hyperedge $e \in E$ is a non-empty subset of $V$. Each hyperedge $e = \{v_1, v_2, \cdots, v_{|e|}\}$ can contain any number of vertices, where $|e|$ denotes the number of vertices in $e$. For each vertex $v \in V$, we use $E_v = \{e_1, e_2, \cdots, e_{|E_v|}\}$ to denote the set of hyperedges that contain $v$. The degree of a vertex $v$, denoted as $d(v)$, is defined as the number of hyperedges containing $v$, i.e., $d(v) = |E_v|$. We use $N_{e_i} = \{e_j \in E \mid e_j \cap e_i \neq \emptyset\}$ to represent all the hyperedges connected to $e_i$, and $N_{v_i} = \{v_j \in V \mid E_{v_j} \cap E_{v_i} \neq \emptyset\}$ to represent all neighbors of $v_i$.
\reviseone{We define a subgraph $H'=(V',E')$ of a hypergraph $H=(V,E)$ as a hypergraph where $V' \subseteq V$ and there exists an injective mapping $\phi: E' \to E$ such that for each $e' \in E'$, we have $e' \subseteq \phi(e')$.}

\begin{definition}[Hyper-vertex Triangle]
\label{def:ho_tri}
Given a hypergraph $H=(V,E)$, and three vertices $v_i, v_j, v_k \in V$ with $E_{v_i} \cap E_{v_j} \neq \emptyset, E_{v_i} \cap E_{v_k} \neq \emptyset$, and $E_{v_j} \cap E_{v_k} \neq \emptyset$, a hyper-vertex triangle ${\bigtriangleup}_{\{v_i, v_j, v_k\}}^{v}$ is a subgraph formed by the three vertices \( v_i, v_j, v_k \) in $H$.
\end{definition}

A hyper-vertex triangle is formed by three interconnected vertices. Based on the number of hyperedges connecting the three vertices $v_i, v_j, v_k$, hyper-vertex triangles can be classified into three patterns: $(i)$ \textit{inner triangles} ${\bigtriangleup}_{inr}$ where three vertices are included in the same hyperedge, i.e., $\{v_i, v_j, v_k\} \subseteq e_1$, $(ii)$ \textit{hybrid triangles} ${\bigtriangleup}_{hyb}$ where three vertices are contained in one hyperedge, while two vertices of them are also contained in another hyperedge, i.e., $\{v_i, v_j, v_k\} \subseteq e_1$ and $\{v_i, v_j\}/\{v_i, v_k\}/\{v_j, v_k\} \subseteq e_2$, and $(iii)$ \textit{outer triangles} ${\bigtriangleup}_{otr}$ where the three vertices are pairwise contained in three different hyperedges, i.e., $\{v_i, v_j\} \subseteq e_1, \{v_i, v_k \}\subseteq e_2$, and $\{v_j, v_k\} \subseteq e_3$.
For example, in Figure~\ref{fig:Hyper-vertex Triangles}, vertices $v_1, v_2, v_3$ are included in $e_1$, forming an inner triangle (Figure~\ref{fig:Hyper-vertex Triangles}(a)). Vertices $v_3, v_4, v_5$ are contained in $e_4$ and $e_5$, forming a hybrid triangle (Figure~\ref{fig:Hyper-vertex Triangles}(b), which also forms an inner triangle since they are contained in $e_5$). Vertices $v_2, v_3, v_5$ appear pairwise in hyperedges $e_1$, $e_2$, and $e_5$, forming an outer triangle (Figure~\ref{fig:Hyper-vertex Triangles}(c)).
The existing work~\cite{zhang2023efficiently} considers only inner and outer triangles, while ignoring hybrid triangles that are equally important, as demonstrated by our experiments in Section~\ref{sec:casestudy}.

\begin{definition}[Hyper-edge Triangle]
\label{def:hy_tri}
Given a hypergraph \( H = (V, E) \) and three hyperedges \( e_i, e_j, e_k \in E \) with \( e_i \cap e_j \neq \emptyset \), \( e_i \cap e_k \neq \emptyset \), and \( e_j \cap e_k \neq \emptyset \), a hyper-edge triangle ${\bigtriangleup}_{\{e_i, e_j, e_k\}}^{e}$ is a subgraph composed of \( e_i, e_j \) and \( e_k \) in \( H \).
\end{definition}

A hyper-edge triangle is formed by three hyperedges, each pair connecting through shared vertices. According to the Figure~\ref{fig:Hyper-edge Triangles}, three hyperedges can partition the vertices into at most seven regions, and depending on the emptiness of these regions, 20 distinct hyper-edge triangle patterns can emerge. Based on the different modes of vertex sharing between any two hyperedges within a hyper-edge triangle, Yin et al.~\cite{yin2025efficient} further categorize these 20 hyper-edge triangle patterns into four distinct classes: $CCC$, $TCC$, $TTC$, and $TTT$, where $T$ represents an intersection, i.e., a pair of hyperedges $e_i \cap e_j \neq \emptyset$ and $|e_i|, |e_j| > |e_i \cap e_j|$, and $C$ represents an inclusion, i.e., $e_i \subset e_j$ or $e_j \subset e_i$. Each $C/T$ indicates that a pair of hyperedges shares vertices through an inclusion/intersection.
For example, in Figure~\ref{fig:Hyper-edge Triangles}, ${\bigtriangleup}_{\{e_2, e_6, e_7\}}^{e}$ belongs to the CCC class (Figure~\ref{fig:Hyper-edge Triangles}(a)); ${\bigtriangleup}_{\{e_2, e_3, e_7\}}^{e}$ belongs to the TCC class (Figure~\ref{fig:Hyper-edge Triangles}(b)); ${\bigtriangleup}_{\{e_3, e_6, e_7\}}^{e}$ belongs to the TTC class (Figure~\ref{fig:Hyper-edge Triangles}(c)); ${\bigtriangleup}_{\{e_1, e_2, e_5\}}^{e}$ belongs to the TTT class (Figure~\ref{fig:Hyper-edge Triangles}(d)).  



\begin{definition}[Hypergraph Stream]
\label{def:hg_stream}
A hypergraph stream $\Pi$ is a sequence of edges:
\[
\Pi = \left( e^{(1)},\ e^{(2)},\ \dots,\ e^{(t)},\ \dots \right)
\]
where each $e^{(i)}$ represents a hyperedge that contains vertices $v_1^{(i)}, v_2^{(i)} \cdots v_{|e^{(i)}|}^{(i)}$, arriving at time $i$. 
\end{definition}

\stitle{Problem Statement.} In this paper, we study the problem
of triangle counting in hypergraph streams. Specifically, given a hypergraph stream $\Pi = (V, E) = ( e^{(1)},\ e^{(2)},\ \dots,\ e^{(t)})$, our goal is to maintain unbiased estimates with low variance of hyper-vertex triangle counts (\textit{inner}, \textit{hybrid}, and \textit{outer} triangles) as well as all patterns of hyper-edge triangle counts under available memory $M$.


Our work focuses on identifying and counting all types of hyper-vertex triangles and four representative classes of hyper-edge triangles, $CCC$, $TCC$, $TTC$, and $TTT$, while it can be easily extended to any specific pattern.
To simplify notation, the superscript $(t)$ may be omitted when the context is clear. 


\section{Memory-aware Triangle Estimation}
\label{sec:method1}

In this section, we first present an overview of our memory-aware sampling algorithm \al, followed by a theoretical analysis of its accuracy, including proofs of unbiasedness and the variance bound. Finally, we analyze the time and space complexity.

\subsection{\al Algorithm}


To address the challenges in existing algorithms, we introduce \al, a  memory-aware sampling algorithm. Unlike traditional methods that predefine the number of hyperedges to sample, which may risk memory overflow or lead to substantial estimation errors due to underutilized memory, our approach dynamically adjusts the number of sampled hyperedges to ensure efficient utilization of the available memory $M$.
\revisetwo{Here, $M$ refers to the number of vertices included in the sampled hyperedges. This directly reflects the actual memory usage, as each vertex is stored as a 32-bit integer in our implementation.} 
Specifically, for each incoming hyperedge, \al applies reservoir sampling to determine whether it should be included in the sample set. If adding the new hyperedge would exceed the memory constraint, \al iteratively removes hyperedges from the sample set at random until the constraint is satisfied. This strategy ensures that each hyperedge is sampled with equal probability.
After successful insertion, \al updates the count estimates for various types of triangles by computing local intersections with the current sample and adjusting the counts using correction factors based on the current sampling probability.
This approach ensures efficient utilization of all available memory without requiring prior knowledge of the hypergraph stream to predefine the number of sampled hyperedges.


In this section, our algorithm primarily focuses on counting hyper-vertex triangles, including inner triangles, hybrid triangles, and outer triangles. However, our method can be easily extended to estimate hyper-edge triangles, as discussed in detail in Section~\ref{sec:Hyper-edge Triangle Counting}.

\stitle{Algorithm.} The pseudo-code of our \al is shown in Algorithm~\ref{algo:deabc}. The algorithm maintains a sample set $G_s$, current memory usage $M_s$, a counter $m$ for the number of hyperedges observed so far and counters for all triangle types (line~1). 
For each incoming hyperedge $e = (v_1, v_2, \dots, v_{|e|})$, we first increment the hyperedge counter $m$. If $|e| \geq 3$, we compute the exact number of inner triangles using the formula $\binom{|e|}{3}$ (lines~4-5). 
Then, we attempt to insert $e$ into the sample set $G_s$ using the \texttt{SampleHyperedge} function. If the memory usage after insertion remains within the limit $M$, $e$ is directly added to the sample (lines~11-13). Otherwise, $e$ is accepted with probability $|G_s| / m$ via a Bernoulli trial (lines~14-17) and, 
if selected, it replaces a randomly chosen hyperedge. If adding the new hyperedge exceeds the memory constraint, hyperedges in the sample set are iteratively removed uniformly at random until the constraint is
satisfied (lines~18-20).
If the insertion is successful, it proceeds to update other triangle count estimates by examining intersections between the new hyperedge and the existing sampled set $G_s$ via the \texttt{UpdateTriangles} function (lines~6-9).
Note that once $M_s$ reaches the memory constraint for the first time, it will only decrease afterward. This design ensures that the sampling probability  of hyperedges is uniform, as increasing the sample size after saturation would imply reinserting previously discarded hyperedges, which is impossible in the one-pass streaming scenario.

\stitle{Triangle Count Estimation.}
Once a hyperedge is inserted, we update the counts for hybrid triangles and outer triangles. The \texttt{UpdateTriangles} function iterates over each pair and triplet formed between the newly inserted hyperedge and the existing hyperedges in the sample to identify relevant triangle structures. 
For each sampled hyperedge $e_j$, if it shares at least one vertex with $e_i$, they may form a hybrid triangle. The hybrid triangle count is updated using the formula $\frac{(|e_i| + |e_j| - 2I_{ij}) I_{ij}(I_{ij} - 1)}{2} \cdot \theta$, where $I_{ij} = |e_i \cap e_j|$ and $\theta =  \frac{m(m - 1)}{|G_s|(|G_s| - 1)}$ is the correction factor used to ensure unbiased estimation (lines~24-29). 
For each such pair $(e_i, e_j)$, we further examine each $e_k$ ($k > j$) in the sample. If all three hyperedges share pairwise intersections but the three-way intersection is empty, they may form an outer triangle. The outer triangle estimate is incremented by $(I_{ij} - I)(I_{ik} - I)(I_{jk} - I) \cdot \gamma$, where $I = |e_i \cap e_j \cap e_k|$ and $\gamma = \frac{m(m - 1)(m - 2)}{|G_s|(|G_s| - 1)(|G_s| - 2)}$ are correction factors (lines~30-35).

\begin{figure}[t]
    \centering

        \includegraphics[width=0.8\textwidth]{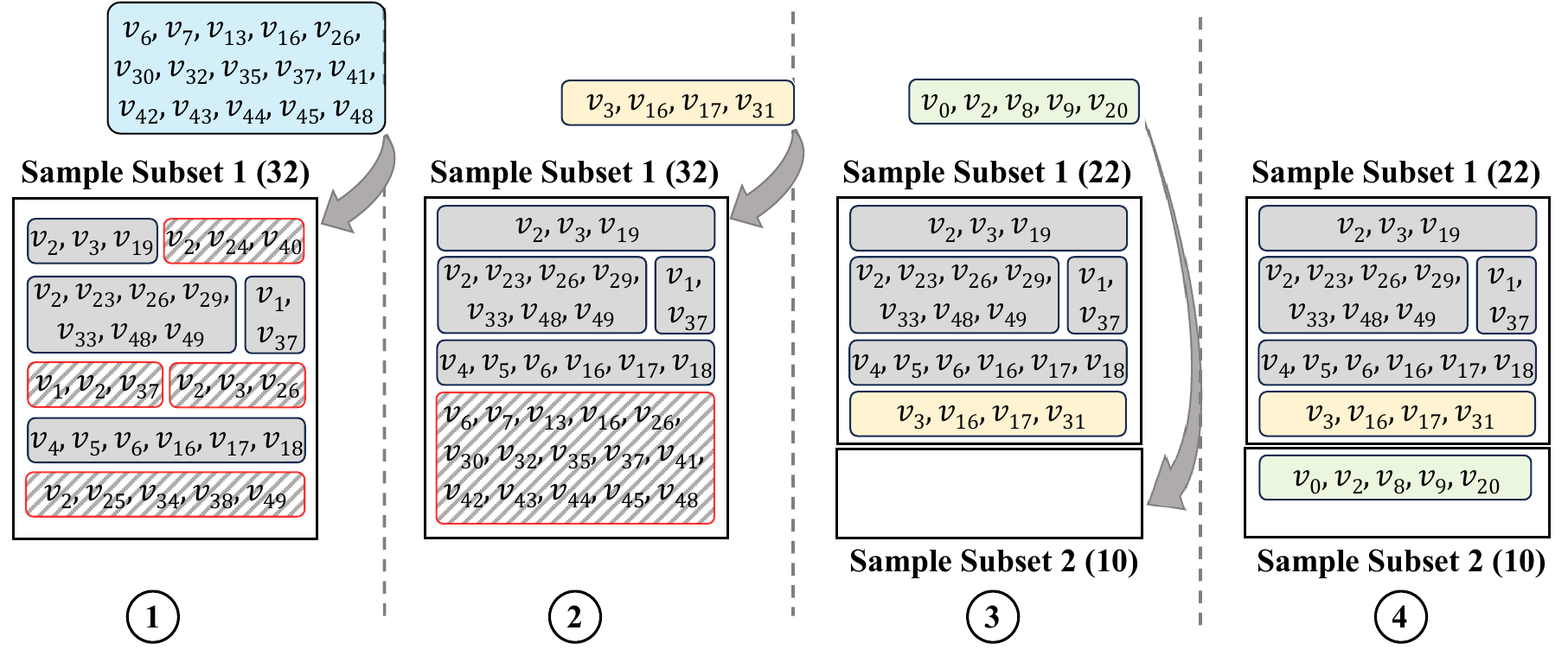}
        \vspace{-1em}
    
    \caption{\revisethree{An Example of Our Algorithms ($M=32, \tau=0.7$)}}

    \label{fig:example_htcount_p}
    \vspace{-1.2em}
\end{figure}

\begin{example}
\revisethree{States~\textcircled{1} and~\textcircled{2} in Figure~\ref{fig:example_htcount_p} illustrate the execution of our \al algorithm with $M=32$. In state~\textcircled{1}, the sample set reaches its memory limit with 8 hyperedges and the triangle counts are $\hat{c}_{{\bigtriangleup}_{inr}} = 59$, $\hat{c}_{{\bigtriangleup}_{hyb}} = 17$, and $\hat{c}_{{\bigtriangleup}_{otr}} = 0$. When a large new hyperedge ($\{v_6, v_7, v_{13}, \cdots, v_{48}\}$) arrives, its inner triangles are counted exactly, increasing $\hat{c}_{{\bigtriangleup}_{inr}}$ to 514. Once sampled, it randomly replaces an existing hyperedge (e.g., $\{v_{2}, v_{24}, v_{40}\}$). If memory usage still exceeds the limit ($44 > M = 32$), additional hyperedges---$\{v_{1}, v_{2}, v_{37}\}$, $\{v_{2}, v_{25}, v_{34}, v_{38}, v_{49}\}$, and $\{v_{2}, v_{3}, v_{26}\}$---are also removed at random until the sample set fits within the constraint. In state~\textcircled{2}, the newly inserted hyperedge forms additional triangles, updating the counts $\hat{c}_{{\bigtriangleup}_{hyb}} = 17 + 42\cdot\theta = 2570.4$ and $\hat{c}_{{\bigtriangleup}_{otr}} = 0 + 0\cdot\gamma = 0$, using correction factors $\theta=3.6$ and $\gamma=8.4$.}
\end{example}

\begin{algorithm}[h]
\small
    \DontPrintSemicolon
    \caption{\al}
    \label{algo:deabc}
    \KwIn{The hypergraph stream $\Pi$ and maximum memory size $M$}
    \KwOut{The estimated number of hyper-vertex triangles $\hat{c}_{{\bigtriangleup}_{inr}}$, $\hat{c}_{{\bigtriangleup}_{otr}}$ and $\hat{c}_{{\bigtriangleup}_{hyb}}$.
    }
    \SetKwComment{comment}{$\triangleright$ }{}

    $G_s \gets \emptyset$; $M_s \gets 0$; $m \gets 0$; $\hat{c}_{{\bigtriangleup}_{inr}} \gets 0$; $\hat{c}_{{\bigtriangleup}_{otr}} \gets 0$; $\hat{c}_{{\bigtriangleup}_{hyb}} \gets 0$; 

    \For{\textbf{each} hyperedge $e = (v_1, v_2, \cdots, v_{|e|}) \in \Pi$}{
    $m \gets m + 1$;\\

         \If{$|e| \ge 3$}{
            $\hat{c}_{{\bigtriangleup}_{inr}} \gets \hat{c}_{{\bigtriangleup}_{inr}} + \frac{|e|(|e|-1)(|e|-2)}{6}$;
         }
        \If{$\textbf{\Sedges($e, m, G_s, M_s, M$)}$}{

            \textbf{if} $|G_s| = m$ \textbf{then} $\theta \gets 1.0$; $\gamma \gets 1.0$;\\
            \textbf{else} $\theta \gets \frac{m(m-1)}{|G_s|(|G_s|-1)}$; $\gamma \gets \frac{m(m-1)(m-2)}{|G_s|(|G_s|-1)(|G_s|-2)}$;\\

            \textbf{\FTriangles($e, \theta, \gamma, G_s$)};\\
        }

    }
    \Fn{\Sedges{$e, m, G_s, M_s, M$}}{
        \If{$M_s + |e| \le M \land |G_s| = m-1$}{
           $G_s \gets G_s \cup \{e\}$; $M_s \gets M_s + |e|$;\\
           \Return true;\\
        }
        \ElseIf{$Bernoulli(\frac{|G_s|}{m}) = 1$}{
            $del \gets  DiscreteUniform(0, |G_s|)$;\\
            $M_s \gets M_s - |G_s[del]|$; $G_s \gets G_s \setminus \{G_s[del]\}$; \\
            $M_s \gets M_s + |e|$; $G_s \gets G_s \cup \{e\}$; \\

            \While{$M_s > M$}{
                $del \gets  DiscreteUniform(0, |G_s|)$;\\
                $M_s \gets M_s - |G_s[del]|$; $G_s \gets G_s \setminus \{G_s[del]\}$; \\
            }
            \Return true;\\
        }
        \Return false;\\
    }
    \Fn{\FTriangles{$e_i, \theta, \gamma, G_s$}}{
        \For{\textbf{each} $e_j \in G_s$}{
            $I_{ij} \gets |e_i \cap e_j|$;\\
            \textbf{if} $I_{ij} = 0$  \textbf{then continue};\\
            \If{$\theta < 0$}{
                $\theta \gets \frac{1}{Pr(e_i, e_j)}$; \comment*{Applied to Algorithm~\ref{algo:deabc2}}
            }
            $\hat{c}_{{\bigtriangleup}_{hyb}} \gets \hat{c}_{{\bigtriangleup}_{hyb}} + \frac{(|e_i| + |e_j| - 2 I_{ij}) I_{ij} (I_{ij} - 1)}{2} \theta$;\\
                \For{\textbf{each} $e_k (k>j) \in G_s$}{
                    $I_{jk} \gets |e_j \cap e_k|$; $I_{ik} \gets |e_i \cap e_k|$; $I \gets |e_i \cap e_j \cap e_k|$; \\
                    \textbf{if} $I_{jk} = 0 \vee I_{ik} = 0$  \textbf{then continue};\\
                    \If{$\gamma < 0$}{
                        $\gamma \gets \frac{1}{\Pr(e_i, e_j, e_k)}$; \comment*{Applied to Algorithm~\ref{algo:deabc2}}
                    }
                    $\hat{c}_{{\bigtriangleup}_{otr}} \gets \hat{c}_{{\bigtriangleup}_{otr}} + (I_{ij} - I)(I_{ik} - I)(I_{jk} - I) \gamma$;\\
                 
                }  
        }
    }
\end{algorithm}

\vspace{-0.3cm}
\subsection{Accuracy Analysis}
\label{sec:accuracy_analysis_in_sec_3}

We now present a detailed theoretical analysis demonstrating that the algorithm (Algorithm~\ref{algo:deabc}) produces unbiased triangle count estimates with low variance. 

\subsubsection{Unbiasedness}

The unbiasedness of our algorithm follows from the fact that each hyperedge in the stream is sampled with equal probability. This is formalized in the following lemma.

\begin{lemma}
\label{the:equal_probability}
In Algorithm~\ref{algo:deabc}, each hyperedge in the hypergraph stream has an equal probability of being sampled up to any time $t$, given by $\frac{|G_s^{(t)}|}{m^{(t)}}$, where $|G_s^{(t)}|$ and $m^{(t)}$ denote the number of sampled and observed hyperedges up to time $t$, respectively.
\end{lemma}

\begin{proof}


\reviseone{
When the sample set is not yet full or only a single hyperedge is replaced, the probability that each hyperedge is sampled can be directly established as $\frac{|G_s^{(t)}|}{m^{(t)}}$ by the classical theory of reservoir sampling.
If multiple replacements are needed to meet the memory constraint, the process is repeated. Assuming $k$ hyperedges are removed, the final probability remains:
\[
  \Pr(e_s \text{ remains}) = \frac{|G_s^{(t-1)}|}{m^{(t)}} \cdot \frac{|G_s^{(t-1)}| - 1}{|G_s^{(t-1)}|} \cdots \frac{|G_s^{(t-1)}| - k}{|G_s^{(t-1)}| - k + 1} = \frac{|G_s^{(t)}|}{m^{(t)}}
\]
}
\end{proof}

\begin{theorem}
\label{the:unbiasedness}
Algorithm~\ref{algo:deabc} provides an unbiased estimate of hyper-vertex triangle count. Specifically, 
$\mathbb{E}[\hat{c}_{{\bigtriangleup}_{inr}}] = c_{{\bigtriangleup}_{inr}}$, $\mathbb{E}[\hat{c}_{{\bigtriangleup}_{hyb}}] = c_{{\bigtriangleup}_{hyb}}$, $\mathbb{E}[\hat{c}_{{\bigtriangleup}_{otr}}] = c_{{\bigtriangleup}_{otr}}$, 
where $ \hat{c}_{{\bigtriangleup}_{inr}}, \hat{c}_{{\bigtriangleup}_{hyb}}, \hat{c}_{{\bigtriangleup}_{otr}}$ 
are the triangle count estimate produced by \al at any time \( t \) and $ c_{{\bigtriangleup}_{inr}}, c_{{\bigtriangleup}_{hyb}}, $ $c_{{\bigtriangleup}_{otr}} $
is the true count.
\end{theorem}

\begin{proof}
\revise{
For inner triangles, the estimator $\hat{c}_{{\bigtriangleup}_{inr}}$ is exactly equal to the true count, since these are directly counted when each hyperedge arrives: 
$
\mathbb{E}[\hat{c}_{{\bigtriangleup}_{inr}}] = c_{{\bigtriangleup}_{inr}}
$.
}

\reviseone{
For hybrid and outer triangles, the unbiasedness relies on whether the probability of a triangle being discovered and counted can be accurately determined at the time it is detected, so that the estimate can be properly corrected. According to Lemma~\ref{the:equal_probability}, each hyperedge has an equal probability $\frac{|G_s|}{m}$ of being sampled, which depends only on the current state of the sample set. Therefore, the probability that a hybrid triangle is counted is $\Pr(\bigtriangleup_{hyb}) = \frac{|G_s|(|G_s|-1)}{m(m-1)}$, and in Algorithm~\ref{algo:deabc}, we use its inverse $\theta = \frac{m(m-1)}{|G_s|(|G_s|-1)}$ as its correction factor. We define the random variable $X_{\bigtriangleup_{hyb}}$ as the contribution of each hybrid triangle; thus,
$
\mathbb{E}[X_{{\bigtriangleup}_{hyb}}]
= \Pr({{\bigtriangleup}_{hyb}}) \times \theta
+ (1-\Pr({{\bigtriangleup}_{hyb}})) \times 0
=1
$.
}

\reviseone{
Therefore, \( \mathbb{E}[\hat{c}_{{\bigtriangleup}_{hyb}}] = \sum_{{\bigtriangleup}_{hyb} \in H} \mathbb{E}[X_{{\bigtriangleup}_{hyb}}] = c_{{\bigtriangleup}_{hyb}} \), which proves that \al provides an unbiased estimate of the hybrid triangle count. 
The same logic applies to outer triangles and the correction factor is
$\gamma = \frac{m(m-1)(m-2)}{|G_s|(|G_s|-1)(|G_s|-2)}$.
}
\end{proof}

\subsubsection{Variance} We now analyze the variance of hyper-vertex triangle count estimates provided by \al.

\begin{theorem}
\label{the:variance}
The variance of hyper-vertex triangle count estimates in Algorithm~\ref{algo:deabc} is bounded as follows:
{\small
\[
\mathrm{Var}[\hat{c}_{{\bigtriangleup}_{inr}}] = 0 ;
\]
\[
\mathrm{Var}[\hat{c}_{{\bigtriangleup}_{hyb}}] \leq (2 c_{{\bigtriangleup}_{hyb}}^2 - c_{{\bigtriangleup}_{hyb}})\frac{m(m-1)}{|G_s|(|G_s|-1)} - c_{{\bigtriangleup}_{hyb}}^2;
\]
\[
\mathrm{Var}[\hat{c}_{{\bigtriangleup}_{otr}}] \leq (2 c_{{\bigtriangleup}_{otr}}^2 - c_{{\bigtriangleup}_{otr}})\frac{m(m-1)(m-2)}{|G_s|(|G_s|-1)(|G_s|-2)} - c_{{\bigtriangleup}_{otr}}^2.
\]
}
\end{theorem}

\begin{proof} 
\reviseone{
For any triangle type $\bigtriangleup$, the variance of its count estimate $\hat{c}_\bigtriangleup$ is given by:
{\small
\begin{align}
\label{eq:var}
\mathrm{Var}[\hat{c}_\bigtriangleup] &= \mathbb{E}[\hat{c}_{\bigtriangleup}^2] - (\mathbb{E}[\hat{c}_{\bigtriangleup}])^2 = 
= \sum_i \mathbb{E}[X_i^2] + \sum_{i \neq j}\mathbb{E}[X_i X_j] - c_{\bigtriangleup}^2 \notag
\end{align}
}
where $\mathbb{E}[\hat{c}_\bigtriangleup] = c_\bigtriangleup$, based on Theorem~\ref{the:unbiasedness}.
}

For inner triangles, each is counted exactly over all hyperedges. Hence $\hat{c}_{\bigtriangleup_{inr}} = c_{\bigtriangleup_{inr}}$ and 
$
\mathrm{Var}[\hat{c}_{\bigtriangleup_{inr}}] = 0.
$

We now consider hybrid triangles. Let random variable $X_i$ denote the contribution of the $i$-th hybrid triangle to the overall count estimate. As defined in Equation~\ref{hybrid_triangle_random_variable}, The expectation of the square is:
$
\mathbb{E}[X_i^2] 
= 1 \cdot p + \theta^2 \cdot (1 - p) \cdot \Pr(\bigtriangleup_{hyb}) 
= p + \theta (1 - p)
$,
where $p = \Pr(T_i \le T_M)$.

\reviseone{
Next, we consider the joint term \( \mathbb{E}[X_i X_j] \). Unlike in traditional graphs, where two triangles can share at most one edge, the situation in hypergraphs is significantly more complex. For instance, two distinct hybrid triangles in a hypergraph can share up to two hyperedges, while outer triangles or other hyper-edge triangles can even share as many as three hyperedges. This richer set of possible overlaps greatly complicates the variance analysis of hypergraph triangle counting algorithms. Considering hybrid triangles, we distinguish three overlap cases between triangles $i$ and $j$: $(i)$ No shared hyperedges; $(ii)$ One shared hyperedge; $(iii)$ Two shared hyperedges. The probability that both $i$ and $j$ are counted is:
}
\vspace{-0.05cm}
{\small
\begin{align}
P_{c1} = 
\begin{cases}
\frac{|G_s|(|G_s|-1)(|G_s|-2)(|G_s|-3)}{m(m-1)(m-2)(m-3)} & case (i) \\
\frac{|G_s|(|G_s|-1)(|G_s|-2)}{m(m-1)(m-2)} & case (ii) \\
\frac{|G_s|(|G_s|-1)}{m(m-1)} & case (iii) \\
0 & \text{otherwise}
\end{cases}
\end{align}
}
\vspace{-0.1cm}

Let $P^{(k)}$ denote the joint probability of both triangles being counted under case $k$ ($k = i, ii, iii$). Then
$
\mathbb{E}[X_i X_j] \le P^{(iii)} \cdot \theta^2 = \frac{m(m-1)}{|G_s|(|G_s|-1)}.
$

Summing up, the variance of $\hat{c}_{\bigtriangleup_{hyb}}$ is:
{\small
\begin{align*}
\mathrm{Var}[\hat{c}_{\bigtriangleup_{hyb}}] 
&\le c_{\bigtriangleup_{hyb}} \cdot \left(p + \theta (1 - p)\right) 
+ 2 c_{\bigtriangleup_{hyb}} (c_{\bigtriangleup_{hyb}} - 1) \cdot \theta - c_{\bigtriangleup_{hyb}}^2 \\
&= (2c_{\bigtriangleup_{hyb}}^2 - c_{\bigtriangleup_{hyb}}) \cdot \theta - c_{\bigtriangleup_{hyb}}^2.
\end{align*}
}

For outer triangles, the analysis follows the same structure, but involves three hyperedges and a correction factor $\gamma = \frac{m(m-1)(m-2)}{|G_s|(|G_s|-1)(|G_s|-2)}$.

The squared expectation becomes:
$
\mathbb{E}[X_i^2] = p + \gamma(1 - p),
$
and the joint term $\mathbb{E}[X_i X_j] \le \gamma$ under the worst-case overlap (three shared hyperedges).
Hence, the variance is bounded by
$
\mathrm{Var}[\hat{c}_{\bigtriangleup_{otr}}] 
\le (2c_{\bigtriangleup_{otr}}^2 - c_{\bigtriangleup_{otr}}) \cdot \gamma - c_{\bigtriangleup_{otr}}^2
$.
\end{proof}




\subsection{Complexity Analysis}

\begin{theorem}
\label{the:time1}
     Algorithm~\ref{algo:deabc} takes $O(m^{(t)} + (|G_s^{(t)}|_{max} + |G_s^{(t)}|_{max} \cdot \ln{\frac{m^{(t)}+1}{|G_s^{(t)}|}})\cdot M^2)$  time to process $t$ elements in the input hypergraph stream, where $|G_s^{(t)}|_{max}$ is the maximum number of hyperedges ever held in the sample space throughout the algorithm.

\end{theorem}

\begin{proof}
Whenever a new hyperedge $e$ arrives, computing its internal triangle count takes $O(1)$ time (line 5). Next, we determine whether $e$ should be inserted into the sample set $G_s$ and carry out the corresponding insertion or replacement operation at an additional cost of $O(1)$ (lines 6-8). Thus, processing all incoming hyperedges at the end of $t$ totals $O(m^{(t)})$ time.
Each time a hyperedge $e$ is successfully inserted into $G_s$, we update the triangle count by checking the intersections between \(e\) and every existing hyperedge in \(G_s\). Iterating over all hyperedges \(e_i\) in \(G_s\) takes \(O(M)\) time, since \(M\) bounds the total number of vertices in the sample. For every \(e_i\) intersecting \(e\), we further check whether there exists some \(e_j \in G_s\) with \(e_i \cap e_j \neq \varnothing\) to form outer triangles or hyper-edge triangles. Consequently, each update requires \(O(M^2)\) time.
When the sample set is not yet full, each edge is accepted with probability 1. Otherwise, it is accepted with probability $\frac{|G_s^{(t)}|}{m^{(t)}+1}$. Consequently, the total number of inserting edges is $|G_s^{(t)}| + \sum_{i=|G_s^{(t)}|}^{m^{(t)}} \frac{|G_s^{(t)}|}{i+1} \approx |G_s^{(t)}| + |G_s^{(t)}| \cdot \ln{\frac{m^{(t)}+1}{|G_s^{(t)}|}}$ based on the approximation formula for harmonic numbers. Since the number of hyperedges in the sample space changes dynamically over time, we take its maximum value $|G_s^{(t)}|_{max}$. Overall, the time cost of Algorithm~\ref{algo:deabc} is $O(m^{(t)} + (|G_s^{(t)}|_{max} + |G_s^{(t)}|_{max} \cdot \ln{\frac{m^{(t)}+1}{|G_s^{(t)}|}})\cdot M^2)$.
\end{proof}

\begin{theorem}
\label{the:space1}
    Algorithm~\ref{algo:deabc} has a space complexity of $O(M)$, where $M$ is the maximum memory size.
\end{theorem}

\begin{proof}
\revisetwo{Algorithm~\ref{algo:deabc} maintains a sample set $G_s$ containing hyperedges whose total size is dynamically controlled to strictly remains within the memory limit $M$. When inserting a new hyperedge causes the total size to exceed $M$, existing hyperedges are removed until the constraint is satisfied (Algorithm~\ref{algo:deabc}, lines 18--20). In addition, our algorithm also uses a small amount of auxiliary space for temporary computations (such as storing intersection results in \texttt{UpdateTriangles} and variables like $\theta$, $\gamma$). However, these auxiliary data structures are at most proportional to the size of a single hyperedge or its intersections, and in practice are much smaller than $M$. Thus, the space complexity is $\mathcal{O}(M)$.}
\end{proof}

\section{Partition-based Triangle Estimation}
\label{sec:method2}

Although \al provides unbiased triangle count estimates with theoretical variance guarantees under memory constraints, it suffers from limited memory efficiency, particularly when hyperedge sizes vary significantly. 
Specifically, once the available memory is saturated, \al maintains feasibility by evicting existing hyperedges from the sample set upon the arrival of a newly sampled one. If the sampled hyperedge is large, it may displace multiple smaller ones, reducing the diversity and representativeness of the sample. Should this large hyperedge be removed later during sampling, the previously evicted hyperedges cannot be recovered, inevitably resulting in wasted memory and degraded estimation accuracy.
\mbox{This phenomenon is also confirmed in our experiment Exp-2.}

To overcome these limitations, we propose a partition-based triangle estimation algorithm \alp. We first present the algorithm details, then provide theoretical analysis, including unbiasedness, the variance bound, and complexity, and discuss the differences from \al.


 


\subsection{\alp Algorithm}

The main idea of \alp is to dynamically partition unused memory into multiple subsets, each independently applying the same hyperedge sampling strategy as \al. When incoming hyperedges arrive, \alp first evaluates the overall memory utilization. If it falls below a predefined threshold~$\tau$, the remaining memory is divided into additional independent sample subsets. Each incoming hyperedge is then routed to one of these subsets according to a weighted discrete distribution, where the weight of each subset is proportional to its current memory allocation. This mechanism enables more fine-grained memory management and enhances the robustness of triangle estimation, particularly under skewed hyperedge-size distributions.

\stitle{Algorithm.} The pseudo-code of \alp is shown in Algorithm~\ref{algo:deabc2}. We initialize up to $N$ empty sample subsets $G_s[1,\dots,N]$, their memory usage $M_s$, hyperedge counters $m$, and a memory allocation vector $M'$, where all memory is initially assigned to the first subset (line~1). 
When an incoming hyperedge arrives, we first check if the number of subsets $\ell$ is less than the maximum $N$ and if the memory utilization is below the threshold $\tau$. If both conditions are met, we add a new subset and update $M'$ (lines~2–7).
To avoid the increased costs from over-fragmentation, we set an upper bound $N$ on the number of sample subsets.
In practice, we find that setting $N$=$10$ achieves a good balance between adaptivity and stability.
Then we select a subset $G_s[p]$ to insert the current hyperedge $e$. If the sampling probability of the last subset is lower than the average of previous subsets, we continue using it and use the flag $canExtend$ to temporarily disable further partitioning (lines~8–10). Otherwise, we assign $e$ to a subset using a weighted random sampling strategy, where the weight is proportional to each subset’s memory allocation (lines~10–12). Once a subset $G_s[p]$ is selected, we increment its hyperedge counter $m[p]$ (line~13). If the hyperedge size $|e| \ge 3$, the exact number of inner triangles is computed and added to the estimator using the combinatorial formula (lines~14-15). We then try to insert $e$ into the selected subset $G_s[p]$ via the \texttt{SampleHyperedge} function. If the insertion is successful, we invoke \texttt{UpdateTriangles} to update triangle estimates (lines~16–17).

\stitle{Triangle Count Estimation.} Since subsets are maintained independently, the joint sampling probability for hyperedges depends on which subsets they belong to. The following lemma gives the sampling probability for each hyperedge.

\begin{lemma}
\label{the:equal_probability2}
In each sampled subset $G_s[i] \in G_s$, the probability that each hyperedge is sampled is equal and depends only on the state of the subset itself, i.e., $\Pr(e_i \text{ is sampled}) = \frac{|G_s[i]|}{m[i]}$ ($1 \le i \le \ell$) where $e_i$ represents the hyperedge assigned to the sampled subset $G_s[i]$.
\end{lemma}

\begin{proof}
\reviseone{
When a hyperedge $e$ arrives, it is assigned to a subset using a weighted discrete distribution~$WeightedDiscrete()$. This assignment is independent for each hyperedge. According to Lemma~\ref{the:equal_probability}, each hyperedge $e$ has an equal probability of being sampled up to any time $t$, given by $\frac{|G_s[i]^{(t)}|}{m[i]^{(t)}}$.
}
\end{proof}


Therefore, the probability that both hyperedges $e_i$ and $e_j$ are sampled is:
{\small
\begin{align}
\label{}
\small
\Pr(e_i, e_j)=
\begin{cases}
\frac{|G_s[x]|(|G_s[x]|-1)}{m[x](m[x]-1)} & e_i, e_j \in G_s[x]\\
\frac{|G_s[x]||G_s[y]|}{m[x]m[y]} & e_i \in G_s[x], e_j \in G_s[y] 
\end{cases}
\end{align}
}

And the three-edge sampling probability is similarly defined as:
{\small
\begin{align}
\label{}
\Pr(e_i, e_j, e_k) =
\begin{cases}
\textstyle \frac{|G_s[x]|(|G_s[x]|-1)(|G_s[x]|-2)}{m[x](m[x]-1)(m[x]-2)} & \text{Case 1} \\
\textstyle \frac{|G_s[x]|(|G_s[x]|-1)|G_s[y]|}{m[x](m[x]-1)m[y]} & \text{Case 2} \\
\textstyle \frac{|G_s[x]||G_s[y]||G_s[z]|}{m[x]m[y]m[z]} & \text{Case 3}
\end{cases}
\end{align}
}
where three cases represent distinct subset configurations:
Case~1: all three hyperedges are from the same subset;  
Case~2: two are from the same subset and one from a different one;  
Case~3: each hyperedge is sampled from a different subset.

When estimating triangle counts, we apply a correction factor to each triangle instance based on the inverse of its sampling probability. Specifically, for hybrid triangles, the correction factor $\theta$ is set as $\frac{1}{\Pr(e_i, e_j)}$, where $e_i$ and $e_j$ are the two hyperedges forming the triangle (line 28 in Algorithm~\ref{algo:deabc}). Similarly, for outer triangles and hyper-edge triangles, the correction factor $\gamma$ is computed as $\frac{1}{\Pr(e_i, e_j, e_k)}$, depending on the subset assignments of the three participating hyperedges (line 35 in Algorithm~\ref{algo:deabc}). 

\begin{example}
    \reviseone{Figure~\ref{fig:example_htcount_p} illustrates a step-by-step example of our \alp algorithm with $M=32$ and $\tau=0.7$. Initially, only one sample subset exists, utilizing the full memory budget. Before new subsets are created (states~\textcircled{1} and~\textcircled{2}), \alp operates identically to \al.
As more hyperedges are sampled and the current subset's utilization drops below $\tau$, a new subset is created and memory is split. For example, in state~\textcircled{3}, after sampling $\{v_3,v_{16},v_{17}, v_{31}\}$ and removing $\{v_6, v_7, v_{13}, \cdots, v_{48}\}$, the utilization falls to $0.69 < 0.7$, which triggers the creation of Sample Subset~2.
New hyperedges (e.g., $\{v_{0}, v_{2}, v_{8}, v_{9}, v_{20}\}$) are added to the new subset until its sampling probability matches the average probability of all subsets.
Finally, in state~\textcircled{4}, both subsets sample independently, and triangle counts are continuously updated based on each subset's sampling probability, ensuring unbiased estimates under strict memory constraints.
}
\end{example}


\begin{algorithm}[h]
\small
    \DontPrintSemicolon
    \caption{\alp}
    \label{algo:deabc2}
    \KwIn{The hypergraph stream $\Pi$, maximum memory size $M$, memory utilization threshold $\tau$ and the maximum number of the sampled subset $N$.}
    \KwOut{The estimated number of hyper-vertex triangles $\hat{c}_{{\bigtriangleup}_{inr}}$, $\hat{c}_{{\bigtriangleup}_{otr}}$ and $\hat{c}_{{\bigtriangleup}_{hyb}}$.
    }
    \SetKwComment{comment}{$\triangleright$}{}

    $G_s[1,\cdots,N] \gets [\emptyset,\cdots,\emptyset]$;
    $M_s[1,\cdots,N] \gets [0,\cdots,0]$; 
    $M'[1,\cdots,N] \gets [M,0,\cdots,0]$; 
    $m[1,\cdots,N] \gets [0,\cdots,0]$; 
    $\ell \gets 1$; $canExtend \gets true$;
    $\hat{c}_{{\bigtriangleup}_{inr}} \gets 0$; $\hat{c}_{{\bigtriangleup}_{otr}} \gets 0$; $\hat{c}_{{\bigtriangleup}_{hyb}} \gets 0$;

    \For{\textbf{each} hyperedge $e = (v_1, v_2, \cdots, v_{|e|}) \in \Pi$}{

        \If{$\ell < N \land canExtend = true \land m[\ell] > |G_s[\ell]| \land \frac{\sum_{i=1}^{\ell}M_s[i]}{M} < \tau$}{
            $M'[1,\cdots,\ell] \gets M_s[1,\cdots,\ell]$;\\
            $\ell \gets \ell +1$;\\
            $M'[\ell] = M - \sum_{i=1}^{\ell-1}M'[i]$;\\
            $canExtend \gets false$;
        }

        \If{$\ell = 1 \vee \frac{|G_s[\ell]|}{m[\ell]} < \frac{1}{\ell -1}\sum_{k=1}^{\ell -1}\frac{|G_s[k]|}{m[k]}$}{
            $p \gets \ell$;
        }
        \Else{
            \mbox{$p \gets WeightedDiscrete(\{(i, \Pr[i]) \mid \Pr[i] = \frac{M'[i]}{\sum_{j=1}^\ell M'[j]}\})$;}\\
            $canExtend \gets true$;

        }

        $m[p] \gets m[p] + 1$;\\

        \If{$|e| \ge 3$}{
            $\hat{c}_{{\bigtriangleup}_{inr}} \gets \hat{c}_{{\bigtriangleup}_{inr}} + \frac{|e|(|e|-1)(|e|-2)}{6}$;
         }

        \If{$\textbf{\Sedges($e, m[p], G_s[p], M_s[p], M'[p]$)}$}{


            \textbf{\FTriangles($e,-1,-1, G_s[p]$)};\\
        }
        
    }

\end{algorithm}

\subsection{Accuracy Analysis}

We now present a detailed theoretical analysis demonstrating that the algorithm (Algorithm~\ref{algo:deabc2}) produces unbiased triangle count estimates with low variance. 

\subsubsection{Unbiasedness}



\begin{theorem}
\label{the:unbiasedness2}
The Algorithm~\ref{algo:deabc2} provides an unbiased estimate of three triangle count. Specifically, 
$\mathbb{E}[\hat{c}_{{\bigtriangleup}_{inr}}] = c_{{\bigtriangleup}_{inr}}$, $\mathbb{E}[\hat{c}_{{\bigtriangleup}_{hyb}}] = c_{{\bigtriangleup}_{hyb}}$, $\mathbb{E}[\hat{c}_{{\bigtriangleup}_{otr}}] = c_{{\bigtriangleup}_{otr}}$, $\mathbb{E}$
where $ \hat{c}_{{\bigtriangleup}_{inr}}, \hat{c}_{{\bigtriangleup}_{hyb}}, \hat{c}_{{\bigtriangleup}_{otr}}$ 
are the triangle count estimate produced by \alp at any time \( t \) and $ c_{{\bigtriangleup}_{inr}}, c_{{\bigtriangleup}_{hyb}}, c_{{\bigtriangleup}_{otr}}$
is the true count.
\end{theorem}

\begin{proof}
Similar to Algorithm~\ref{algo:deabc}, inner triangles are directly counted based on the number of vertices each incoming hyperedge contains, that is,  \( \hat{c}_{{\bigtriangleup}_{inr}} = c_{{\bigtriangleup}_{inr}} \). Therefore, the estimate of the inner triangle count is unbiased.

We also define the random variable \( X_{{\bigtriangleup}_{hyb}} \), representing the contribution of each
hybrid triangle. According to Algorithm~\ref{algo:deabc2}, when a hybrid triangle ${\bigtriangleup}_{hyb}$ is found, we compensate its count by correction factor $\lambda = \frac{1}{\Pr(e_i,e_j)}$, where $e_i$ and $e_j$ are the hyperedges that form ${\bigtriangleup}_{hyb}$, i.e.,
{\small
\begin{align}
\label{hybrid_triangle_random_variable}
X_{{\bigtriangleup}_{hyb}}=
\begin{cases}
\frac{1}{\Pr(e_i,e_j)} & {\bigtriangleup}_{hyb} \text{ is counted}  \\
0 & \text{otherwise}
\end{cases}
\end{align}
}

Since the probability of \(\triangle_{hyb}\) being detected is \(\Pr(e_i, e_j)\), the expected value of \(X_{\triangle_{hyb}}\) is given by
$
\mathbb{E}[X_{\triangle_{hyb}}] = \frac{1}{\Pr(e_i, e_j)} \cdot \Pr(e_i, e_j) = 1.
$
Hence, summing over all hybrid triangles gives,
$
\mathbb{E}[\hat{c}_{\triangle_{hyb}}] = \sum_{\triangle_{hyb} \in H} \mathbb{E}[X_{\triangle_{hyb}}] = c_{\triangle_{hyb}}.
$
which proves that Algorithm~\ref{algo:deabc2} provides an unbiased estimate of the hybrid triangle count. 


The same reasoning applies to outer triangles, where the correction factor is $\frac{1}{\Pr(e_i, e_j, e_k)}$ and the corresponding sampling probability is $\Pr(e_i, e_j, e_k)$, ensuring that the estimation is unbiased.
\end{proof}

\subsubsection{Variance}

\label{sec:variance2}
We now analyze the variance of the triangle count estimates produced by \alp. 

\begin{theorem}
The variance of the hyper-vertex triangle count estimate in Algorithm~\ref{algo:deabc2} is bounded as follows:
{\small
\[
\mathrm{Var}[\hat{c}_{{\bigtriangleup}_{inr}}] = 0 ;
\]
\[
\mathrm{Var}[\hat{c}_{{\bigtriangleup}_{hyb}}] \leq (2 c_{{\bigtriangleup}_{hyb}}^2 - c_{{\bigtriangleup}_{hyb}})\Phi_1 - c_{{\bigtriangleup}_{hyb}}^2;
\]
\[
\mathrm{Var}[\hat{c}_{{\bigtriangleup}_{otr}}] \leq \left( 2 c_{{\bigtriangleup}_{otr}}^2 - c_{{\bigtriangleup}_{otr}} \right)\Phi_2  - c_{{\bigtriangleup}_{otr}}^2.
\]
}
where
{\small
\[
\Phi_1 = \max_{x \in [1, \ell]} \frac{m[x](m[x]-1)}{|G_s[x]|(|G_s[x]|-1)},
\]
\[
\Phi_2 = \max_{x \in [1, \ell]} \frac{m[x](m[x]-1)(m[x]-2)}{|G_s[x]|(|G_s[x]|-1)(|G_s[x]|-2)}
\]
}


\end{theorem}

\begin{proof}
For any triangle type $\triangle$, the variance of its count estimate $\hat{c}_\bigtriangleup$ is given by:
{\small
\begin{align}
\label{eq:var}
\mathrm{Var}[\hat{c}_\bigtriangleup]
&= \sum_i \mathbb{E}[X_i^2] + \sum_{i \neq j}\mathbb{E}[X_i X_j] - c_{\bigtriangleup}^2 \notag
\end{align}
}

Inner triangles are also counted exactly in Algorithm~\ref{algo:deabc2}; therefore, $\mathrm{Var}[\hat{c}_{\bigtriangleup_{inr}}] = 0$. For other types of triangles, based on the previous analysis, the probability of each hybrid/outer triangle being observed is $\Pr(e_i, e_j)/\Pr(e_i, e_j, e_k)$. When it is counted, we apply the correction factor, $ \theta = \frac{1}{\Pr(e_i, e_j)}$ or $\gamma = \frac{1}{\Pr(e_i, e_j, e_k)}$, to compensate. 
For any triangle types, let \(\Pr_i\) be the probability that triangle \(i\) is sampled, and let \(X_i = \frac{1}{\Pr_i}\) if triangle \(i\) is detected, and 0 otherwise. Then we calculate:
{\small
\[
\mathbb{E}[X_i^2] = \left( \frac{1}{\Pr_i} \right)^2 \cdot {\text{Pr}_{i}} = \frac{1}{\Pr_i}
\]
\[
\mathbb{E}[X_i X_j] \le \frac{1}{\Pr_i \cdot \Pr_j} \cdot \max(\text{Pr}_i, \text{Pr}_j) = \frac{1}{\min(\Pr_i, \Pr_j)}
\]
}



Assuming that the worst-case sampling probability among all triangles, we further bound:
{\small
\[
\mathrm{Var}[\hat{c}_\triangle]
\le (2 c_\triangle^2 - c_\triangle) \cdot \Phi - c_\triangle^2
\]
}
where \(\Phi = \max_{x \in [1, \ell]} \frac{m[x](m[x]-1)}{|G_s[x]|(|G_s[x]|-1)}\) for hybrid triangles formed through the interaction of two hyperedges, or \(\Phi = \max_{x \in [1, \ell]} \frac{m[x](m[x]-1)(m[x]-2)}{|G_s[x]|(|G_s[x]|-1)(|G_s[x]|-2)}\) for outer triangles formed through the interaction of three hyperedges.
\end{proof}

\stitle{\underline{\revise{Remark.}}}
\reviseone{
Compared to \al, calculating the variance for partition based algorithm \alp is more complex, as each sample subset has its own size and sampling probability. 
This requires considering all possible subset assignments for the hyperedges in a triangle.
Despite this added complexity, \alp achieves a lower variance overall.
Taking the estimation of hybrid triangles as an example, $\Phi$ is given by $\frac{m(m-1)}{|G_s|(|G_s|-1)}$ in \al and $\max_{x \in [1, \ell]} \frac{m[x](m[x]-1)}{|G_s[x]|(|G_s[x]|-1)}$ in \alp. In \al, once the sample size $|G_s|$ becomes saturated, the ratio $m/|G_s|$ continues to increase, resulting in a linear growth of $\Phi$ over time. In contrast, \alp will reset $m[\ell]$ once a new sample subset $G_s[\ell]$ is created, and new hyperedges are only assigned to other subsets when the ratio $m[\ell]/|G_s[\ell]|$ in the latest subset reaches that of previous subsets, which effectively stalls the growth of $\Phi$ in earlier ones. As a result, even the largest $\Phi$ in \alp is typically smaller than that in \al, that is $\max_{x \in [1, \ell]} \frac{m[x](m[x]-1)}{|G_s[x]|(|G_s[x]|-1)} < \frac{m(m-1)}{|G_s|(|G_s|-1)}$. Therefore, the variance in \alp is lower overall.
}


\subsection{Complexity Analysis}

\begin{theorem}
    Algorithm~\ref{algo:deabc2} takes $O(\sum_{j=1}^{\ell}(m[j]^{(t)} + (|G_s[j]^{(t)}|_{max} $ $+ |G_s[j]^{(t)}|_{max} \cdot \ln{\frac{m[j]^{(t)}+1}{|G_s[j]^{(t)}|}})\cdot M^2))$  time to process $t$ elements in the input hypergraph stream, where $|G_s[j]^{(t)}|_{max}$ is the maximum number of hyperedges ever held in the sample subset $j$.
\end{theorem}

\begin{proof}
When processing each incoming hyperedge $e$, the algorithm first checks and possibly creates new subsets or decides which subset the hyperedge should be assigned to. These steps (lines 2-11) require constant time $O(1)$. Since sampling and counting within each sampled subset are performed independently, according to Theorem~\ref{the:time1}, the time complexity for subset $j$ is 
\[
m[j]^{(t)} + (|G_s[j]^{(t)}|_{max} + |G_s[j]^{(t)}|_{max} \cdot \ln{\frac{m[j]^{(t)}+1}{|G_s[j]^{(t)}|}})\cdot M^2)
\]
Therefore, the total time complexity is 
\[
O(\sum_{j=1}^{\ell}(m[j]^{(t)} + (|G_s[j]^{(t)}|_{max} + |G_s[j]^{(t)}|_{max} \cdot \ln{\frac{m[j]^{(t)}+1}{|G_s[j]^{(t)}|}})\cdot M^2))
\]
where $\ell$ denotes the number of sample subsets actually used. 
\end{proof}

\begin{theorem}
    Algorithm~\ref{algo:deabc2} has a space complexity of $O(M)$, where $M$ is the maximum memory size.
\end{theorem}

\begin{proof}
\revisetwo{Algorithm~\ref{algo:deabc2} partitions the total memory $M$ into up to $N$ sample subsets. Each subset only stores hyperedges up to the limit imposed by its current memory allocation \( M'[i] \), and the sum
\(
\sum_{j=1}^{N} M'[j] = \sum_{j=1}^{\ell} M'[j] \leq M
\)
always holds. Thus, the total space used by Algorithm~\ref{algo:deabc2} is bounded by $O(M)$.}
\end{proof}

\stitle{\underline{Remark.}}
\revisetwo{
Considering the time complexity, \al outperforms \alp. \alp incurs an extra step to choose a subset for each incoming hyperedge. 
Whenever overall memory utilization drops below the threshold, \alp creates a new sample subset, leading to the sampling of more hyperedges, i.e., $\sum_{j=1}^{\ell}|G_s[j]|_{max} > |G_s|_{max}$.
Overall, \alp is better suited for scenarios with highly variable hyperedge sizes, as its adaptive multi-sample method ensures better memory utilization and robustness.}


\section{Hyper-edge Triangle Counting}
\label{sec:Hyper-edge Triangle Counting}
In this section, we extend our proposed algorithms, \al and \alp, to estimate hyper-edge triangle counts, focusing on the four representative classes (CCC, TCC, TTC, and TTT) for clarity of presentation.
Our algorithms, however, can easily be applied to all 20 hyper-edge triangle patterns with minimal changes.


\stitle{Update Procedure.} 
Each incoming hyperedge is sampled using the same strategy as in Algorithm~\ref{algo:deabc}.
Unlike hyper-vertex triangles, the classification of hyper-edge triangles relies on pairwise interactions among three hyperedges, categorized as either intersection (T) or inclusion (C). When a hyperedge $e$ is sampled, Algorithm~\ref{algo:deabc} already computes the intersections $I_{ij}, I_{jk}, I_{ik}$ between $e$ and other sampled hyperedges. To identify the interaction type, we check for inclusion relationships (e.g., $I_{ij} = \min(|e_i|, |e_j|)$).
We then increment the corresponding triangle count using a correction factor $\gamma = \frac{1}{\Pr(e_i, e_j, e_k)}$, i.e., $\hat{c}_{\bigtriangleup} \gets \hat{c}_{\bigtriangleup} + \gamma$, 
where ${\bigtriangleup} \in \{CCC, TCC, TTC, TTT\}$ denotes the hyper-edge triangle class.

\stitle{Accuracy Analysis.} Similar to the accuracy analysis presented for hyper-vertex triangles, we first define the random variable,
{\small
\vspace{-0.2em}
\begin{align*}
\label{outer_triangle_random_variable}
X_{{\bigtriangleup}}=
\begin{cases}
\gamma & {\bigtriangleup} \text{ is counted } \\
0 & \text{otherwise}
\end{cases}
\end{align*}
\vspace{-0.2em}
}

Each hyper-edge triangle is sampled in the probability $\frac{1}{\gamma}$ and adjusted by correction factor $\gamma$. Then we have,

\begin{theorem}\label{the:het_al}
By applying \al to estimate hyper-edge triangle counts, we have:
{\small
\vspace{-0.2em}
\[
\mathbb{E}[\hat{c}_{{\bigtriangleup}}] = c_{{\bigtriangleup}}
;~~~
\mathrm{Var}[\hat{c}_{{\bigtriangleup}}] \leq (2 c_{{\bigtriangleup}}^2 - c_{{\bigtriangleup}})\frac{m(m-1)(m-2)}{|G_s|(|G_s|-1)(|G_s|-2)} - c_{{\bigtriangleup}}^2
\]
\vspace{-0.2em}
}
where ${\bigtriangleup} \in \{CCC, TCC, TTC, TTT\}$
\end{theorem}

\begin{theorem}\label{the:het_alp}
By applying \alp to estimate hyper-edge triangle counts, we have:
$
\mathbb{E}[\hat{c}_{{\bigtriangleup}}] = c_{{\bigtriangleup}}
;~~~
\mathrm{Var}[\hat{c}_{{\bigtriangleup}}] \leq \left( 2 c_{{\bigtriangleup}}^2 - c_{{\bigtriangleup}} \right)\Phi  - c_{{\bigtriangleup}}^2
$,
where $
\Phi = \max_{x \in [1, \ell]} \frac{m[x](m[x]-1)(m[x]-2)}{|G_s[x]|(|G_s[x]|-1)(|G_s[x]|-2)}
$ and ${\bigtriangleup} \in \{CCC,$ $ TCC, TTC, TTT\}$.
\end{theorem}

The proofs of Theorem~\ref{the:het_al} and Theorem~\ref{the:het_alp} are similar to those presented for hyper-vertex triangle counting and are therefore omitted here for brevity.



\section{Experimental Evaluation}
\label{sec:ee}

\begin{table}[t]
    \caption{Datasets}
    \label{tab:datasets}
    \begin{tabular}{c|c|c|c|c|c}
    \hline
    Datasets & $|V|$ & $|E|$ & $|e|_{min}$ & $|e|_{max}$ & $|e|_{avg}$  \\ \hline

    MAG           &   80,198     &   51,889   &   2    &  25  &   3.5   \\ \hline
    Walmart           &    88,860     &   69,906   &   2    &   25 &   6.6     \\ \hline
    NDC           &   5,311    &   112,405   &   2    &   25   &   4.3 \\ \hline
    Trivago-clicks            & 172,738       &  233,202   &  2   &   86   &  3.1     \\ \hline
    Congress-bills    &  1,718  &  260,851  & 2   &  400 & 8.7  \\ \hline
    MAG-Geology       &  1,256,385  &  1,590,335  &  2  & 284  &  2.8  \\ \hline
    DBLP           &  1,924,991  &  3,700,067  &  2  &  25 &  3.4  \\ \hline
    Threads-stack   &  2,675,955  & 11,305,343 &  2  &  25 &  2.6 \\ \hline

    \end{tabular}
    
\end{table}

\begin{figure}[]
    \centering


    
    \begin{minipage}{0.24\textwidth}
        \centering
        \includegraphics[width=\textwidth]{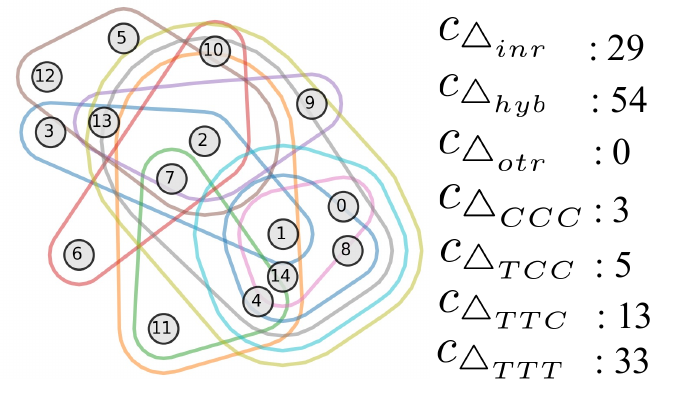}
        \subcaption{\dblp}
    \end{minipage}
    \begin{minipage}{0.24\textwidth}
        \centering
        \includegraphics[width=\textwidth]{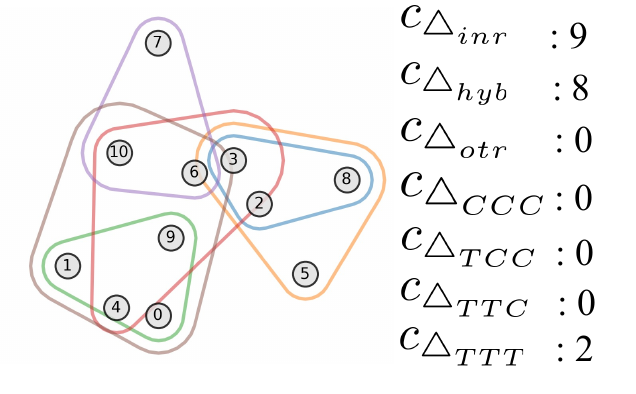}
        \subcaption{\mg}
    \end{minipage}
    \begin{minipage}{0.24\textwidth}
        \centering
        \includegraphics[width=\textwidth]{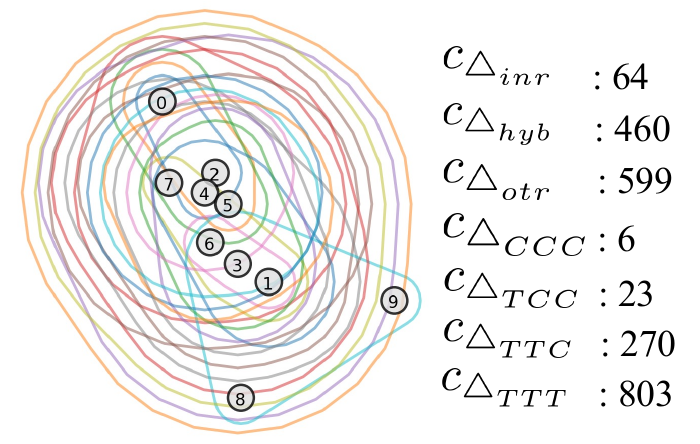}
        \subcaption{\ndc}
    \end{minipage}
    \begin{minipage}{0.24\textwidth}
        \centering
        \includegraphics[width=\textwidth]{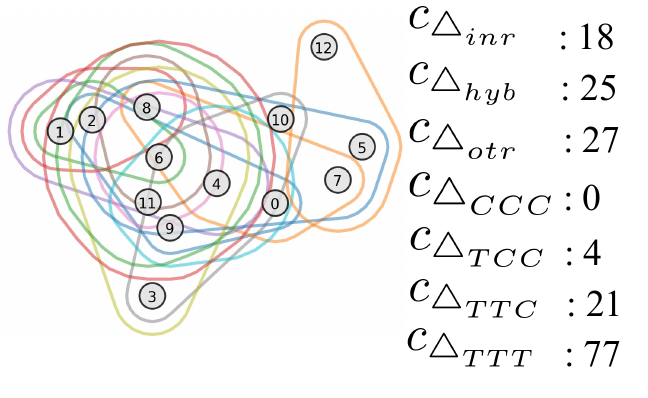}
        \subcaption{\ts}
    \end{minipage}
    \caption{Triangle Counts from Real-world Datasets}
    \label{fig:demo1}
\end{figure}
\begin{figure}
    \centering

    \includegraphics[width=0.59\textwidth]
    {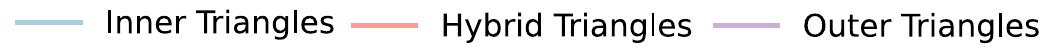}
    
    \begin{minipage}{0.48\textwidth}
        \includegraphics[width=\textwidth]{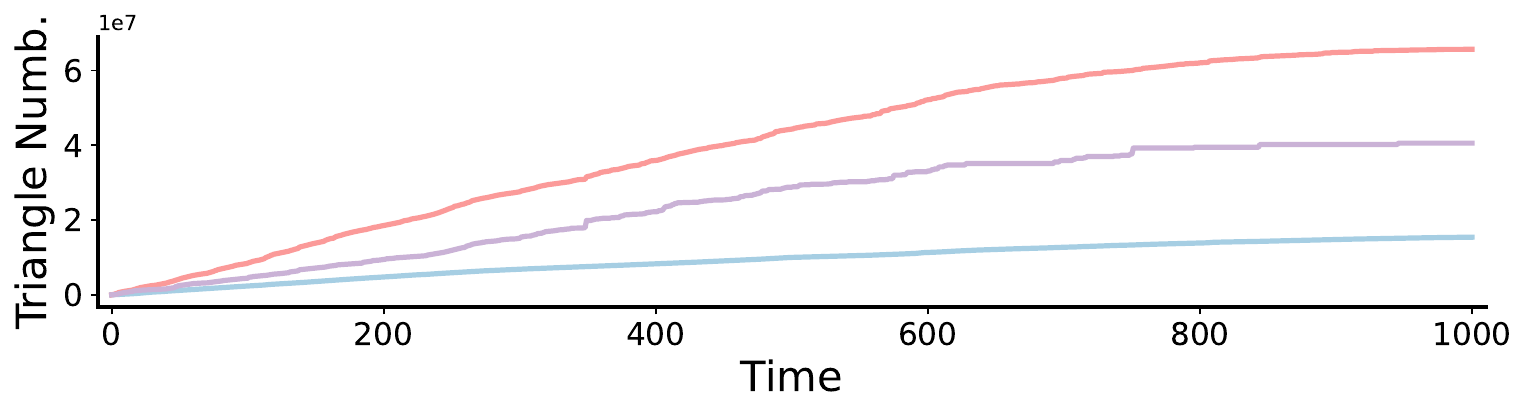}
        \subcaption{\dblp}
    \end{minipage}
    \begin{minipage}{0.48\textwidth}
        \includegraphics[width=\textwidth]{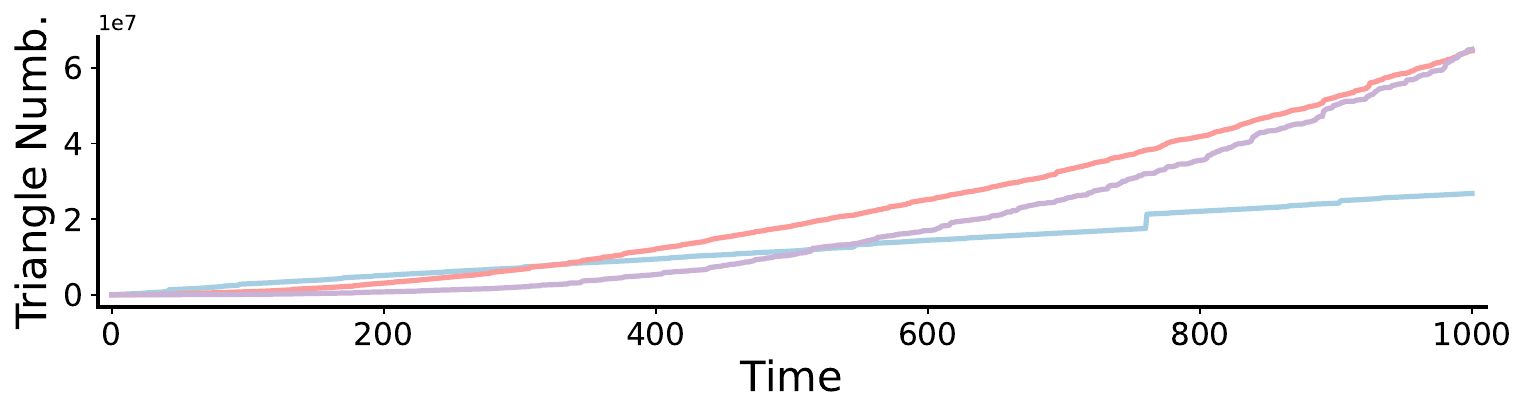}
        \subcaption{\mg}
    \end{minipage}


    \caption{The Number of Triangles over Time}
    \label{fig:case2}
\end{figure}

\begin{figure*}[]
    \centering
    \vspace{-0.8em}
    \includegraphics[width=0.99\textwidth]{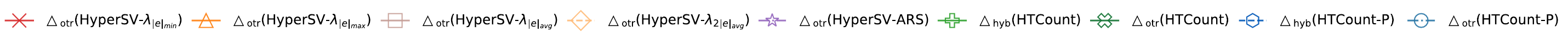}

    \begin{minipage}{0.245\textwidth}
        \centering
        \includegraphics[width=\textwidth]{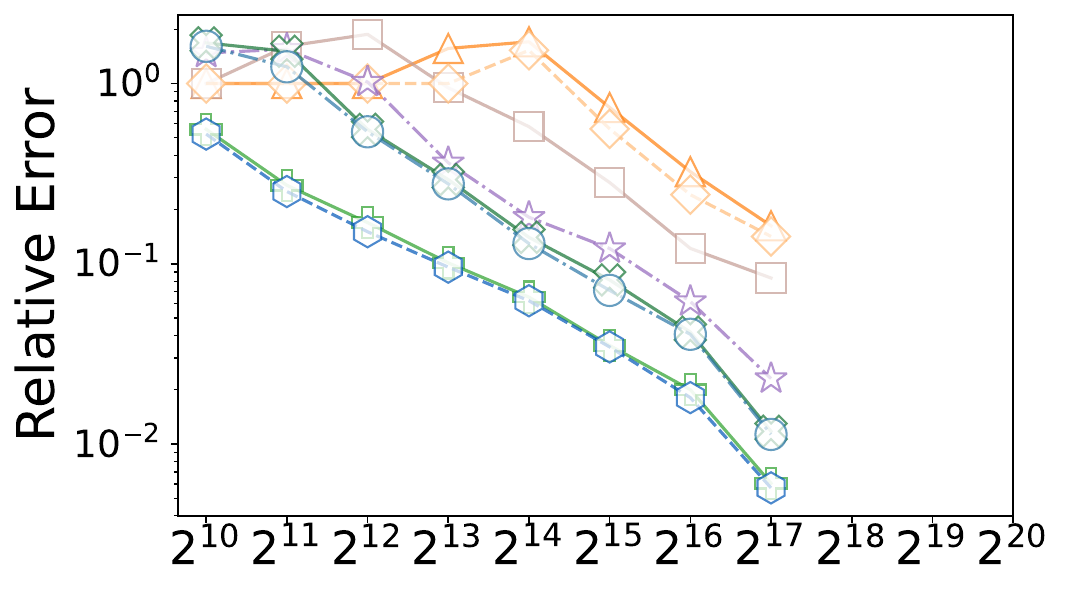}
        \subcaption{\ma (M)}
    \end{minipage}
    \begin{minipage}{0.245\textwidth}
        \centering
        \includegraphics[width=\textwidth]{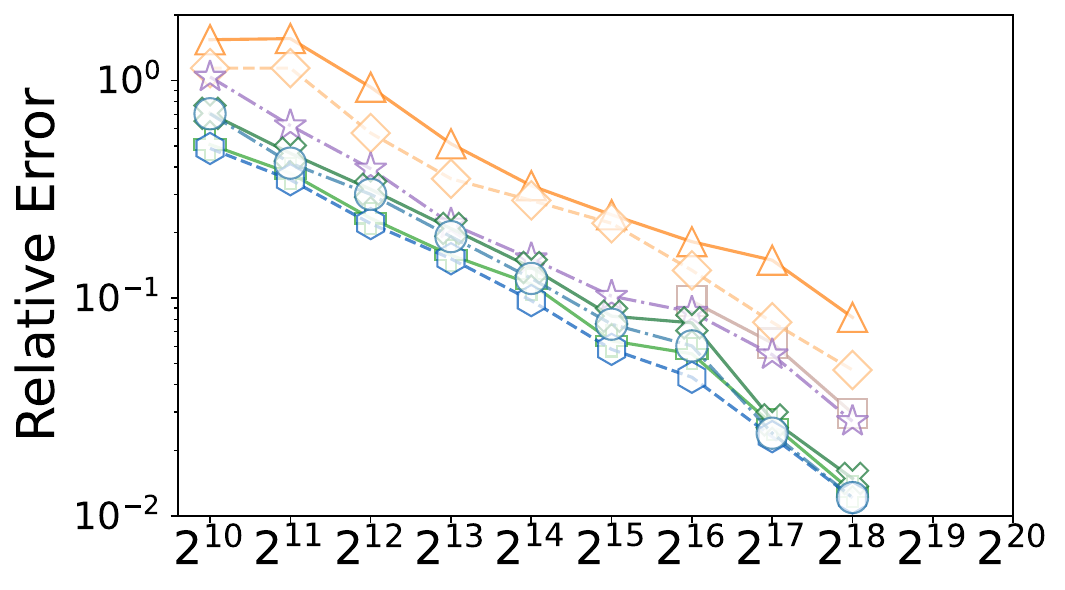}
        \subcaption{\walmart (M)}
    \end{minipage}
    \begin{minipage}{0.245\textwidth}
        \centering
        \includegraphics[width=\textwidth]{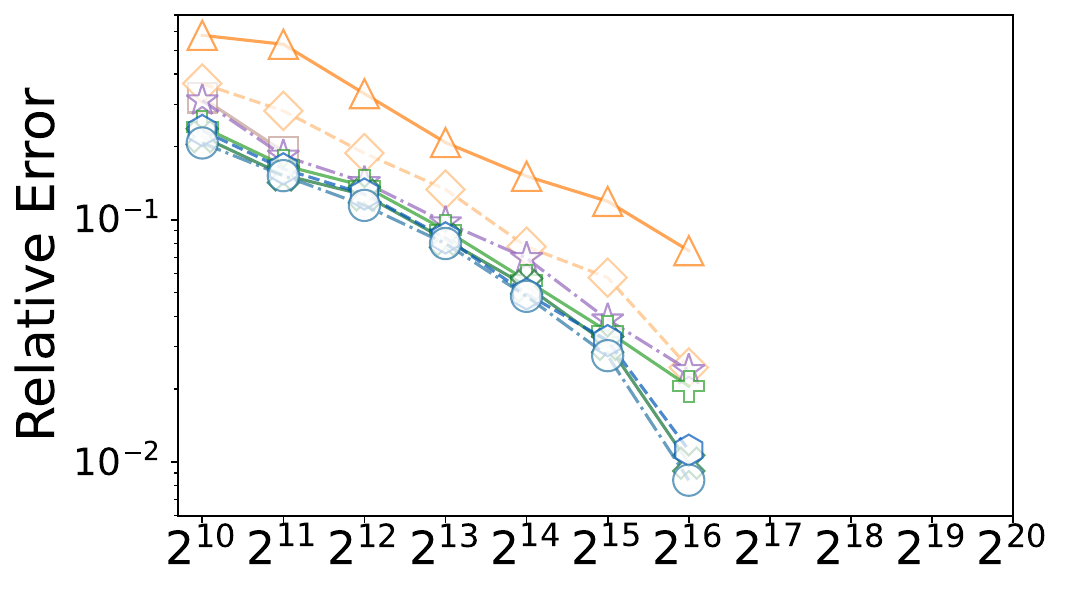}
        \subcaption{\ndc (M)}
    \end{minipage}
    \begin{minipage}{0.245\textwidth}
        \centering
        \includegraphics[width=\textwidth]{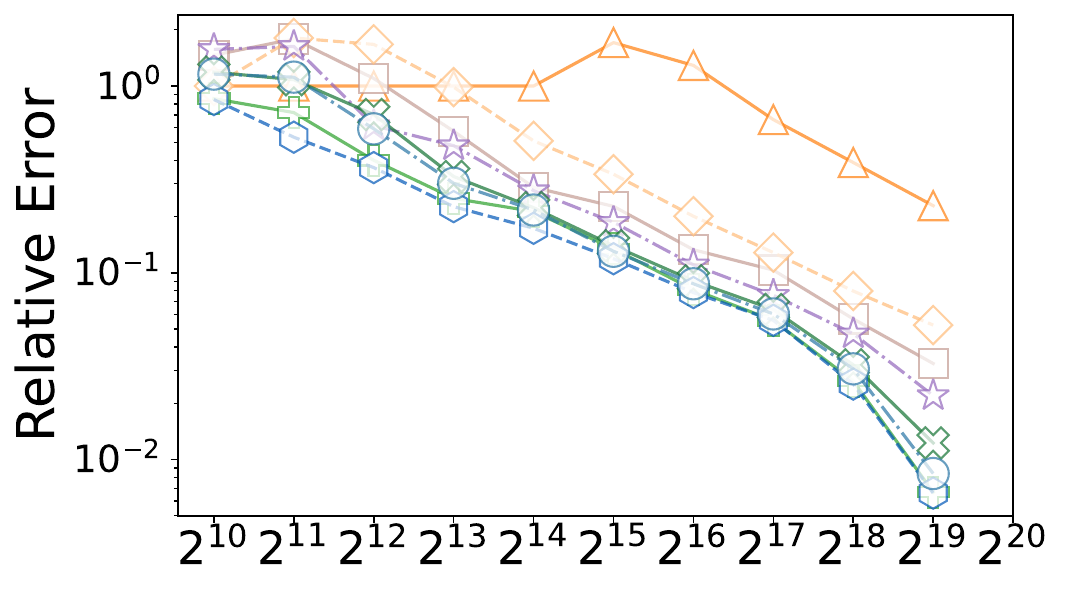}
        \subcaption{\tc (M)}
    \end{minipage}

    \begin{minipage}{0.245\textwidth}
        \centering
        \includegraphics[width=\textwidth]{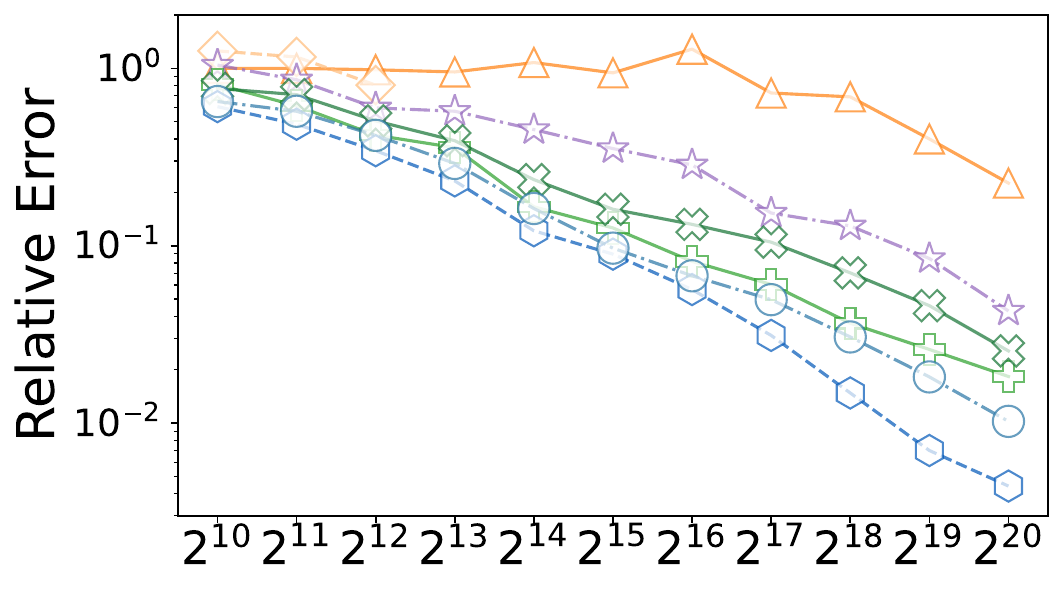}
        \subcaption{\cb (M)}
    \end{minipage}
    \begin{minipage}{0.245\textwidth}
        \centering
        \includegraphics[width=\textwidth]{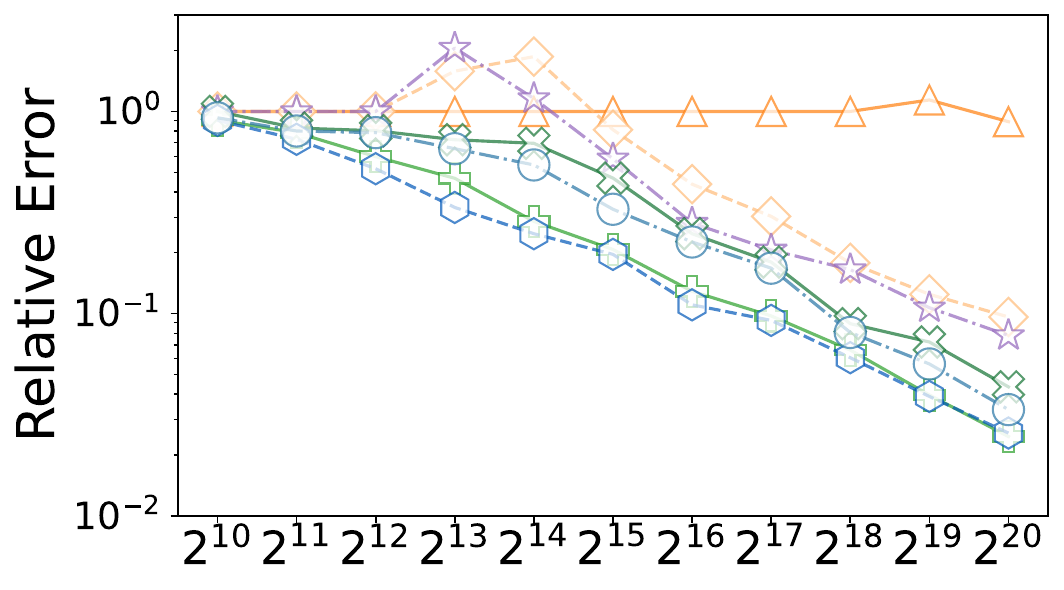}
        \subcaption{\mg (M)}
    \end{minipage}
    \begin{minipage}{0.245\textwidth}
        \centering
        \includegraphics[width=\textwidth]{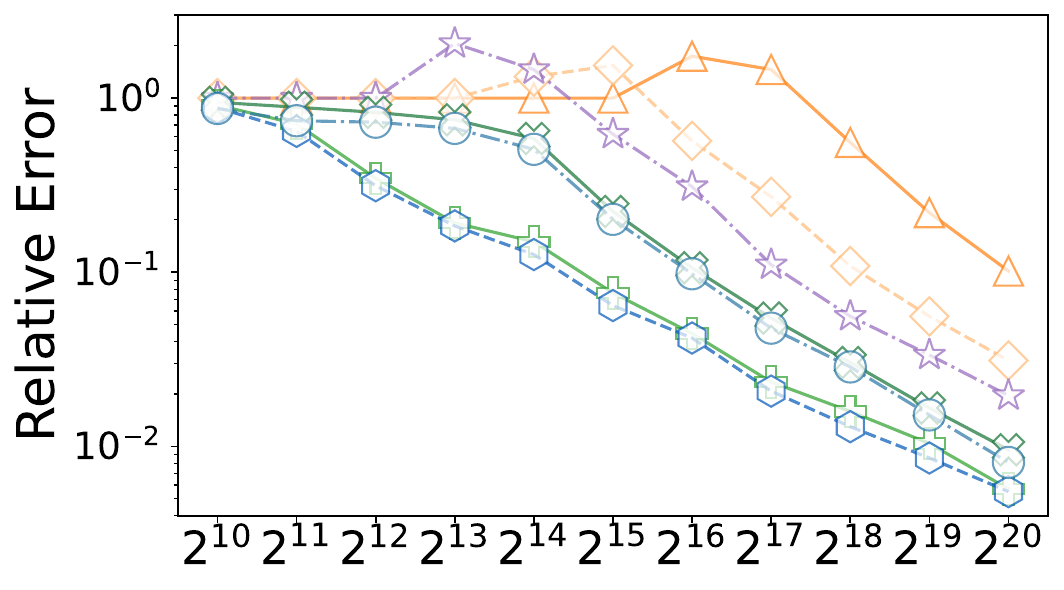}
        \subcaption{\dblp(M)}
    \end{minipage}
    \begin{minipage}{0.245\textwidth}
        \centering
        \includegraphics[width=\textwidth]{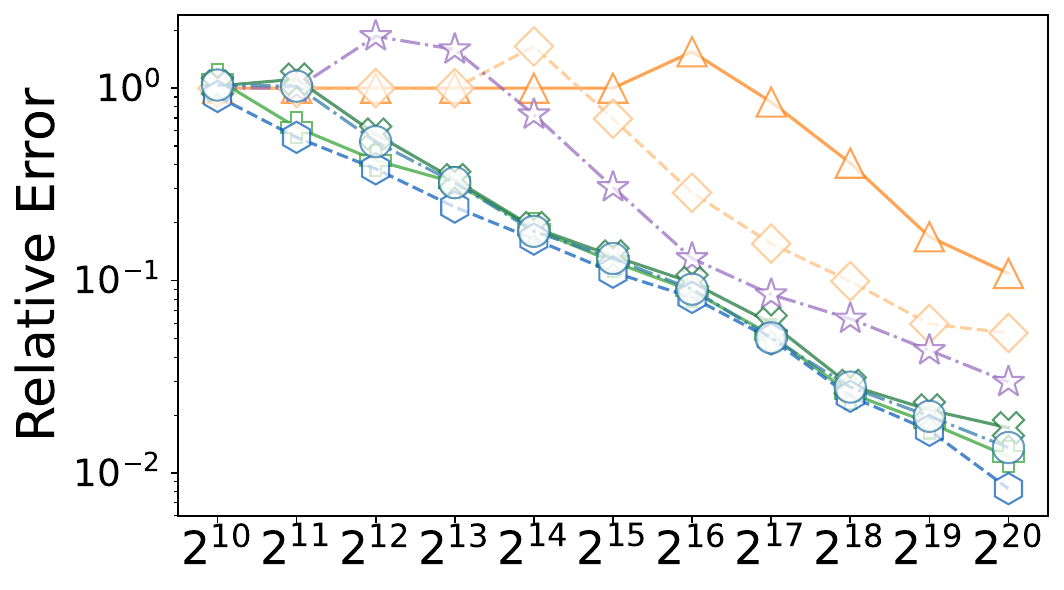}
        \subcaption{\ts (M)}
    \end{minipage}

    \caption{\revisethree{Relative Error of Hyper-vertex Triangle Counting under Different Sample Sizes}}
    \label{fig:exp_accuracy}
\end{figure*}
\begin{figure*}[]
    \centering
    \includegraphics[width=0.99\textwidth]{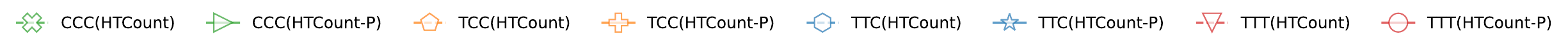}

    \begin{minipage}{0.245\textwidth}
        \centering
        \includegraphics[width=\textwidth]{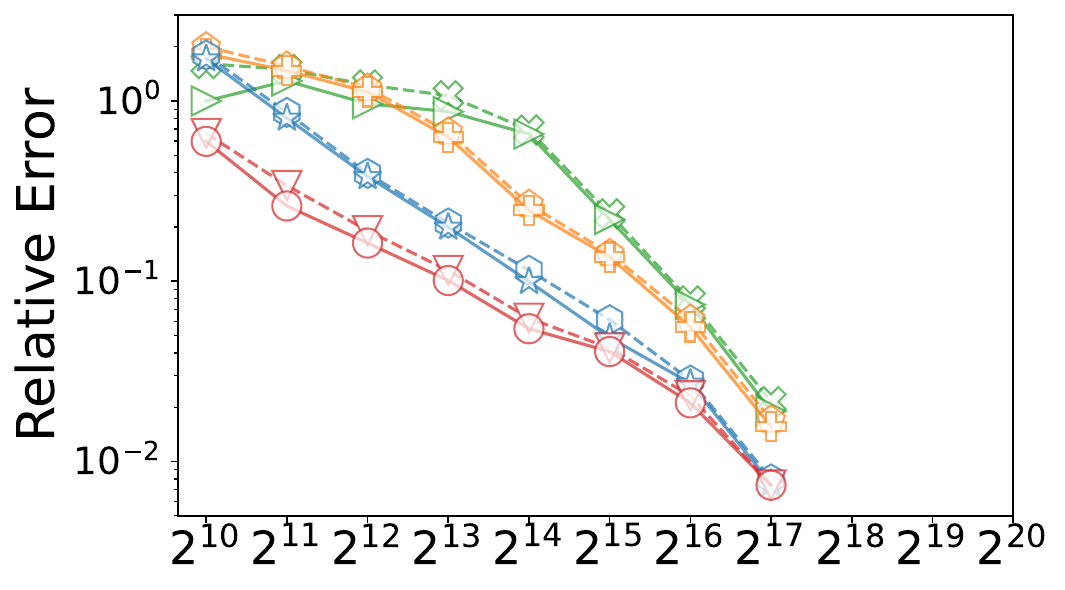}
        \subcaption{\ma (M)}
    \end{minipage}
    \begin{minipage}{0.245\textwidth}
        \centering
        \includegraphics[width=\textwidth]{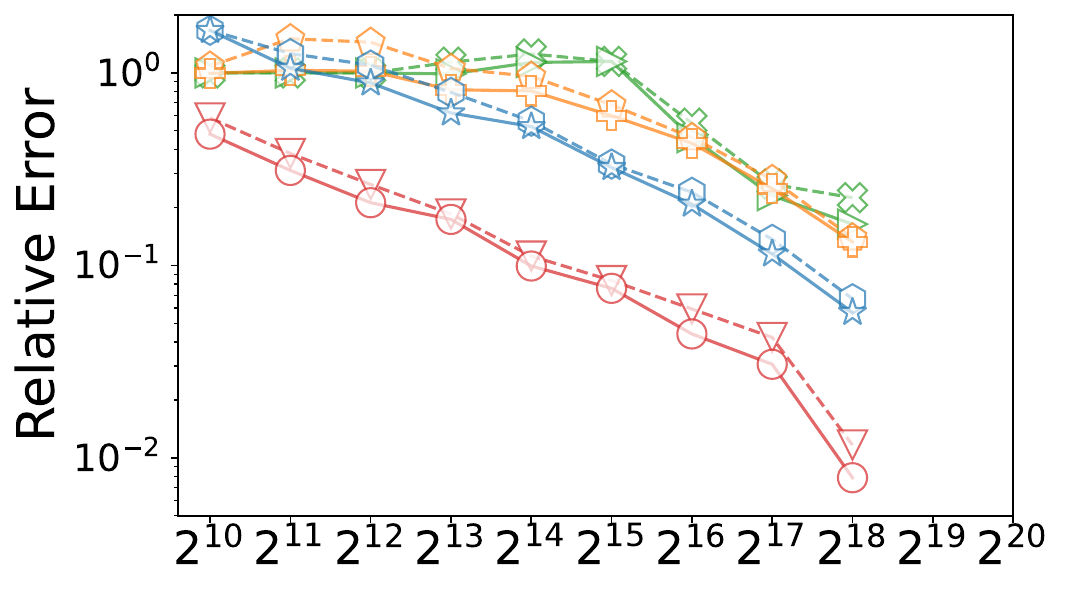}
        \subcaption{\walmart (M)}
    \end{minipage}
    \begin{minipage}{0.245\textwidth}
        \centering
        \includegraphics[width=\textwidth]{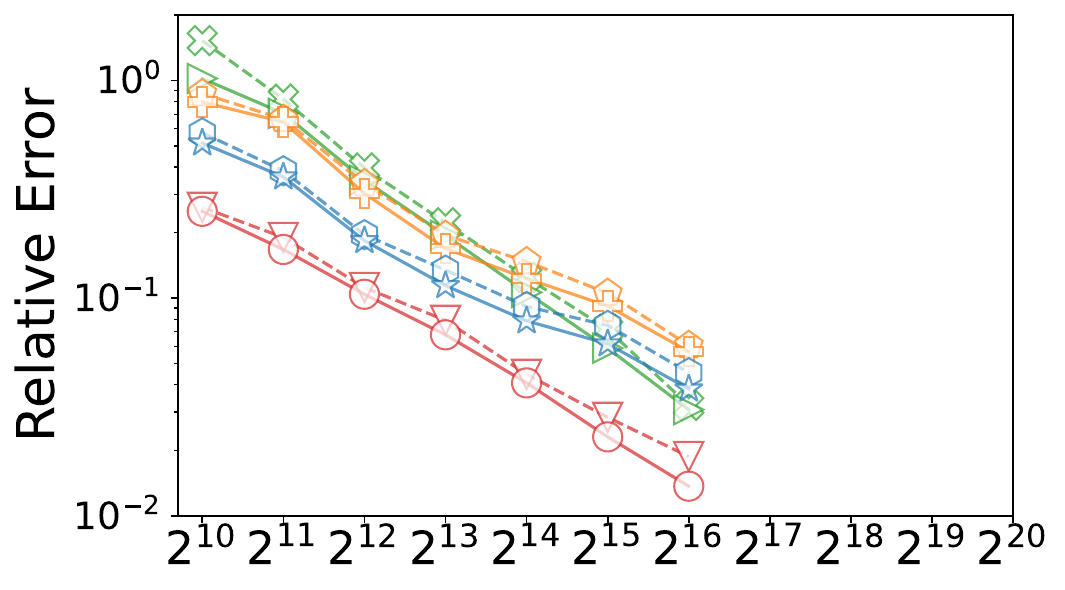}
        \subcaption{\ndc (M)}
    \end{minipage}
    \begin{minipage}{0.245\textwidth}
        \centering
        \includegraphics[width=\textwidth]{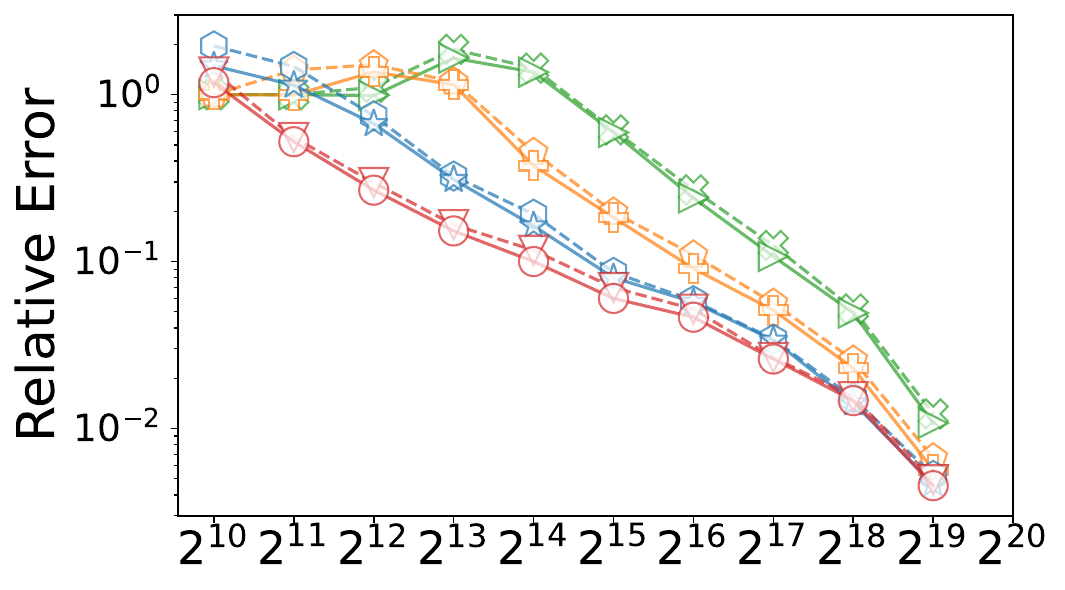}
        \subcaption{\tc (M)}
    \end{minipage}

    \begin{minipage}{0.245\textwidth}
        \centering
        \includegraphics[width=\textwidth]{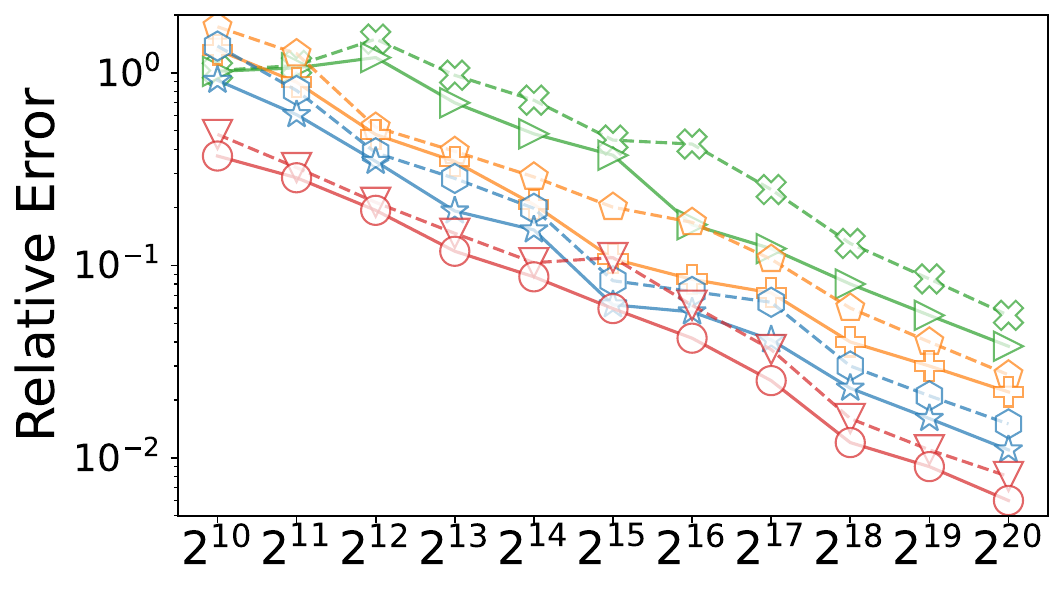}
        \subcaption{\cb (M)}
    \end{minipage}
    \begin{minipage}{0.245\textwidth}
        \centering
        \includegraphics[width=\textwidth]{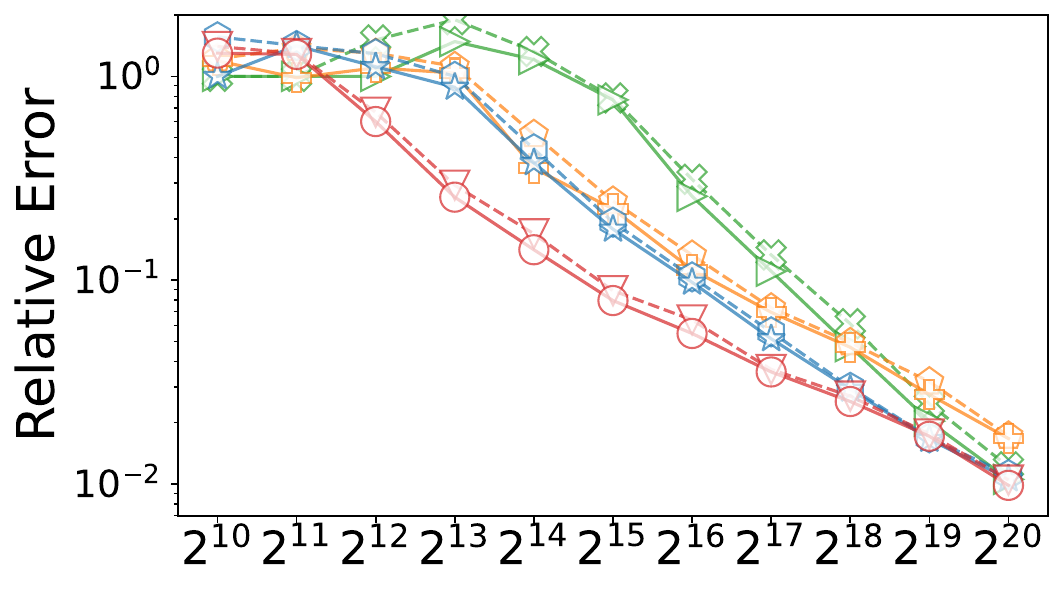}
        \subcaption{\mg (M)}
    \end{minipage}
    \begin{minipage}{0.245\textwidth}
        \centering
        \includegraphics[width=\textwidth]{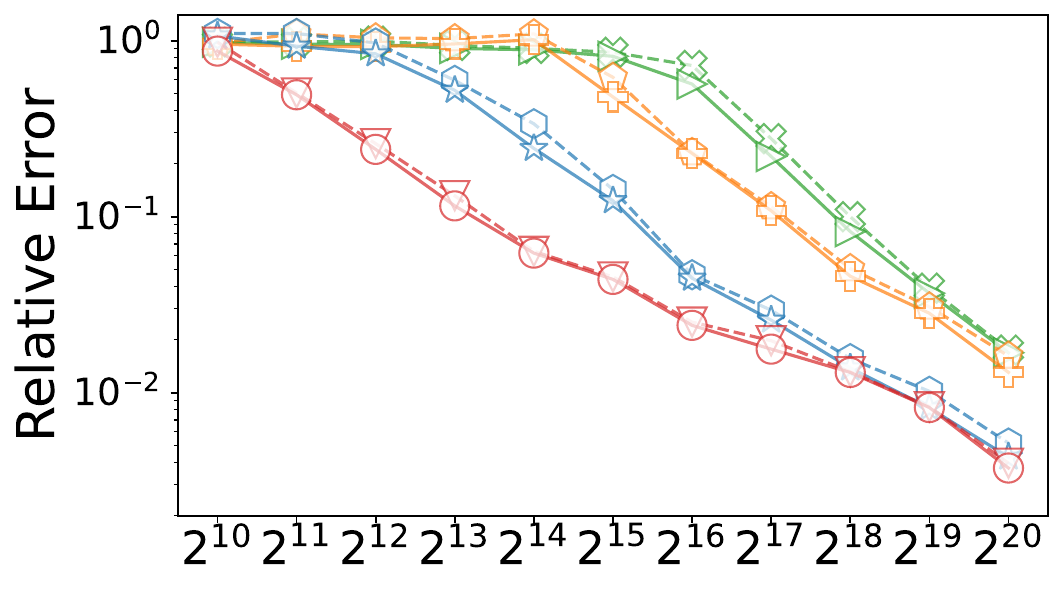}
        \subcaption{\dblp(M)}
    \end{minipage}
    \begin{minipage}{0.245\textwidth}
        \centering
        \includegraphics[width=\textwidth]{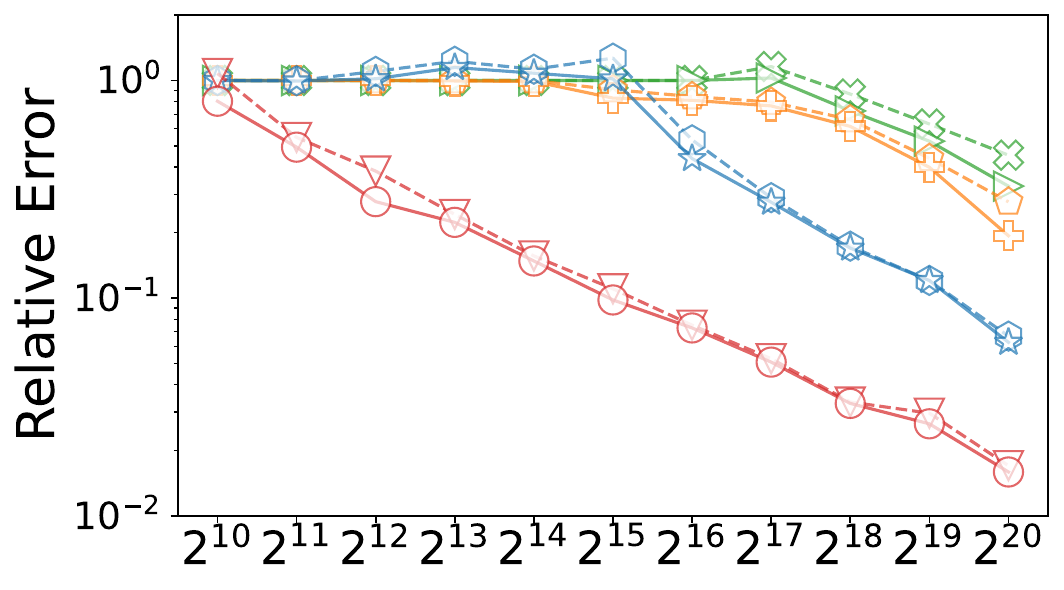}
        \subcaption{\ts (M)}
    \end{minipage}

    \caption{\revisethree{Relative Error of Hyper-edge Triangle Counting under Different Sample Sizes}}
    
    \label{fig:exp_accuracy2}
\end{figure*}

In this section, we evaluate the effectiveness and efficiency of our algorithms, implemented in C++ and compiled with GNU GCC 4.8.5 using the \texttt{-O3} optimization flag. The experiments run on an Intel(R) Xeon(R) Platinum 8373C CPU @ 2.60GHz with 16GB of RAM, and execution time is measured as wall-clock time.


\stitle{Datasets.} 
We use 8 public datasets in our experiments (see Table~\ref{tab:datasets}). \ma is a subset of the Microsoft Academic Graph, with authors as vertices and co-authored papers as hyperedges~\cite{Sinha-2015-MAG}. \walmart represents sets of co-purchased products at Walmart~\cite{Amburg-2020-categorical}. \tc is a hotel hypergraph where vertices are accommodations and hyperedges are sets of accommodations clicked by a user in a single session~\cite{chodrow2021hypergraph}.
Other datasets are from~\cite{benson2018simplicial}. \dblp and \mg are co-authorship networks; \ndc contains sets of substances in drugs; \cb is a network of U.S. congressional bills, with vertices as congresspersons and hyperedges as bill sponsor groups; and \ts represents groups of users $Q\&A$ questions.


\stitle{Algorithms.} We evaluate the performance of our methods, \al and \alp, against the state-of-the-art \hypersv~\cite{zhang2023efficiently} for triangle counting over hypergraph streams.
Since our inter-triangle count is exact and \hypersv does not define hybrid triangles or identify different hyper-edge triangle patterns, we compare only outer triangle counts. All optimizations in~\cite{zhang2023efficiently} are already applied to \hypersv.
\revisethree{We also implement and evaluate the adaptive-size reservoir sampling~\cite{al2007adaptive} variant of \hypersv (denoted as \hypersvars) for fair comparison with our methods.}



\stitle{Evaluation Metrics.} Our evaluation focuses on three primary metrics following~\cite{meng2024counting,zhang2023efficiently,DBLP:conf/icde/PapadiasKPQM24,DBLP:journals/tkdd/SheshboloukiO22}: \textit{Relative Error}, \textit{Throughput} and \textit{Memory Utilization}. \textit{Relative Error} (the lower the better) quantifies the accuracy of an estimate by measuring the normalized difference between the estimated and true triangle counts:. Its formula is as follows:
$
\textit{Relative Error} = \frac{|c_{\bigtriangleup} - \hat{c}_{\bigtriangleup}|}{c_{\bigtriangleup}} 
$.
\reviseone{\textit{Throughput} (the larger the better) measures the amount of data processed by the algorithm per second (KB/s). }
\textit{Memory Utilization} (the larger the better) measures the memory efficiency of the algorithm by calculating the proportion of utilized memory, that is:
$
\text{Memory Utilization} = \frac{M_{\text{utilized}}}{M} \times 100\%
$.

\subsection{Case Study}
\label{sec:casestudy}
Different hypergraph applications show varied interaction patterns, revealed through analyzing the counts of triangles~\cite{10.14778/3407790.3407823,yin2025efficient}.
We present two case studies demonstrating our complete triangle model's advantages over the existing method~\cite{zhang2023efficiently}.


\stitle{Case Study 1: Real-world Implications of Hypergraph Triangle Counting.}
We perform detailed analyses on subgraphs with 10 to 15 vertices from four real-world datasets, as shown in Figure~\ref{fig:demo1}.

By analyzing the distribution of triangles, we observe similar trends in \dblp and \mg: only inner and hybrid triangles exist, with TTT dominating hyper-edge classes. This indicates frequent cross-group collaboration and most collaborations occur between sub-teams within larger research groups---an interaction pattern characteristic of structural fold groups, which facilitates cross-team knowledge integration and are early signals of interdisciplinary convergence~\cite{vedres2010structural,de2015game,wang2010discovering}. 
Detecting these structures reveals early interdisciplinary trends, whereas ignoring hybrid triangles (as in prior models) misleadingly suggests isolated research groups—contradicting Figure~\ref{fig:demo1}(a) and (b). In the \ndc dataset, outer triangles dominate, but significant hybrid triangles and prevalent TTT patterns indicate frequent cross-medication ingredient reuse, revealing combinatorial relationships rather than isolated usage. Similarly, \ts exhibits comparable patterns, demonstrating that incomplete models overlooking hybrid triangles miss critical insights, further validating the necessity of a complete hyper-vertex triangle model.

\stitle{Case Study 2: Tracking Hyper-vertex Triangle Counts in Hypergraph Streams.}
Tracking hypergraph triangle trends reveals collaboration dynamics and emerging research areas, offering insights for science policy and early identification of new fields~\cite{ko2022growth,lee2022mining}. As shown in Figure~\ref{fig:case2}, we examine the trends of hyper-vertex triangles over time in two academic co-authorship networks, \dblp and \mg. 


Hybrid triangles dominate in both \dblp and \mg, reflecting evolving collaboration patterns. In \dblp, hybrid triangles rapidly surpass others, signaling early formation of tight-knit, domain-specific groups. \mg initially shows inner triangle dominance, but hybrid triangles surge after timestep 300, marking interdisciplinary emergence. Outer triangles notably rise after timestep 500, indicating broader cross-team collaboration. This hybrid growth serves as a transitional phase between isolated and cross-team work, providing earlier convergence signals than methods focusing solely on inner/outer triangles.

\vspace{-0.5em}
\subsection{Performance Evaluations}
\vspace{-0.2em}
\label{sec:pe}

\begin{figure}[t]
    \centering

    \includegraphics[width=0.7\textwidth]{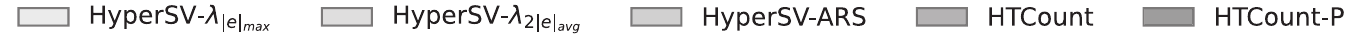}

    
    \begin{minipage}{0.48\textwidth}
        \centering
        \includegraphics[width=\textwidth]{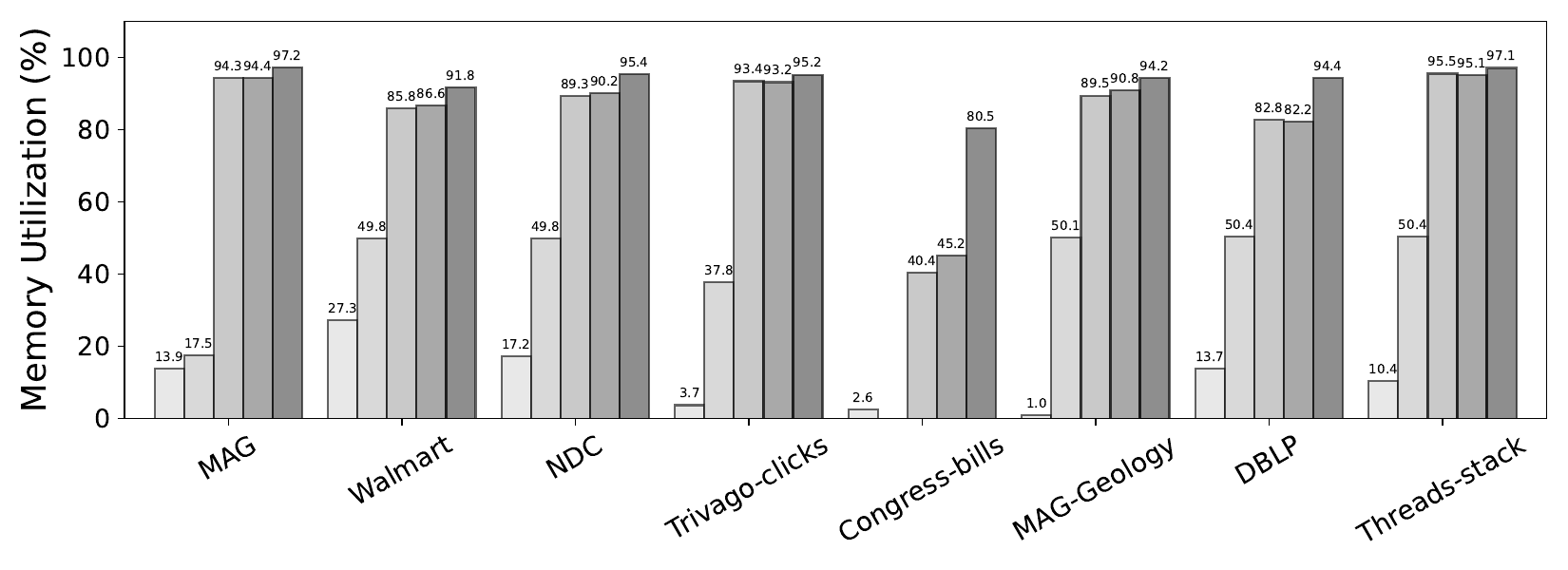}
        \subcaption{Sample Size = 10}
    \end{minipage}
    \hspace{0.015\linewidth}
    \begin{minipage}{0.48\textwidth}
        \centering
        \includegraphics[width=\textwidth]{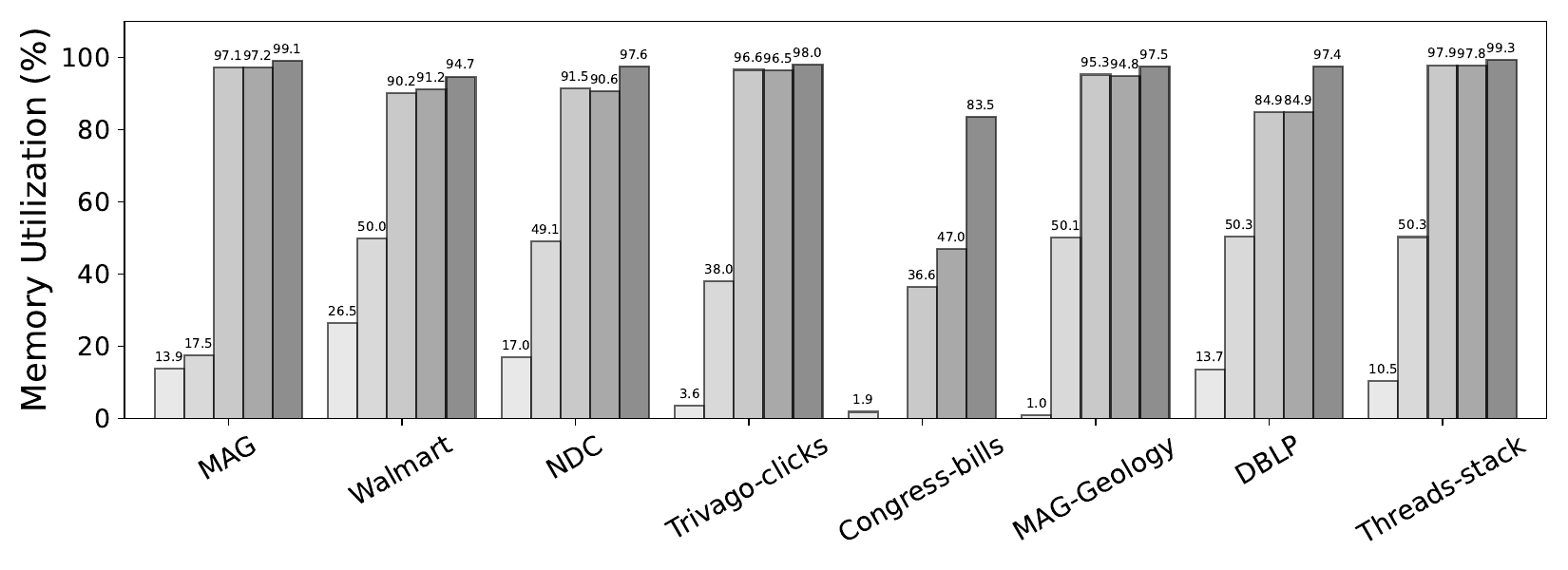}
        \subcaption{Sample Size = 12}
    \end{minipage}

    \caption{\revisethree{Comparison of Memory Utilization}}
    \label{fig:wasted_space}
\end{figure}

    


    


    


    

\begin{figure*}[htbp]
    \centering

    \includegraphics[width=0.7\textwidth]{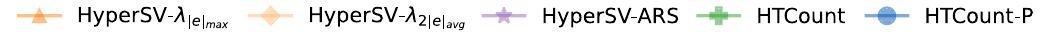}

    \begin{minipage}{0.98\textwidth}
        \centering
        \includegraphics[width=\textwidth]{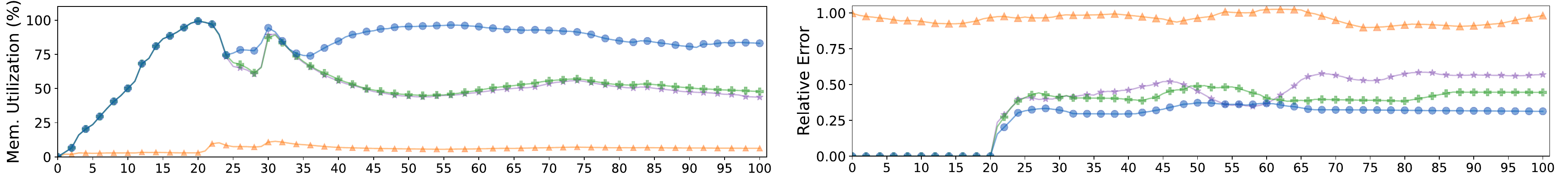}

        \subcaption{\cb}
    \end{minipage}%

    \begin{minipage}{0.98\textwidth}
        \centering
        \includegraphics[width=\textwidth]{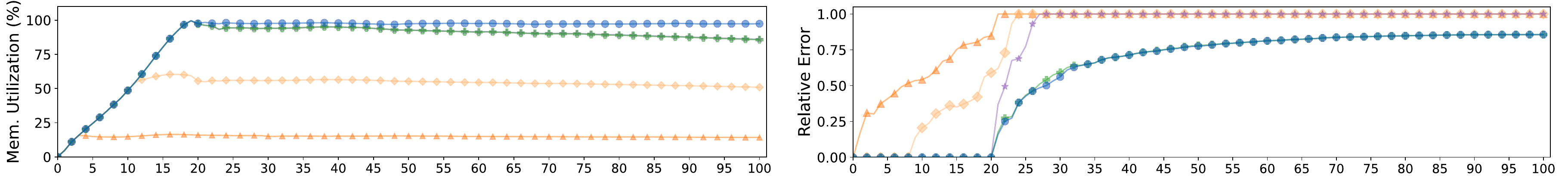}

        \subcaption{\dblp}
    \end{minipage}


    \caption{
        \revisethree{Memory Utilization and Relative Error over Time ($M=2^{12}$) on \cb\ and \dblp}
    }
    \label{fig:mu_over_time}
\end{figure*}

\begin{figure}[t]
    \centering

    \includegraphics[width=0.6\textwidth]{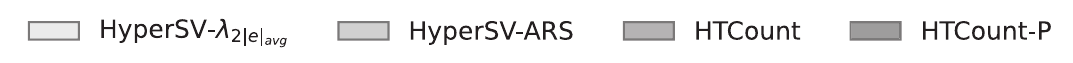}

    
    \begin{minipage}{0.7\textwidth}
        \centering
        \includegraphics[width=0.95\textwidth]{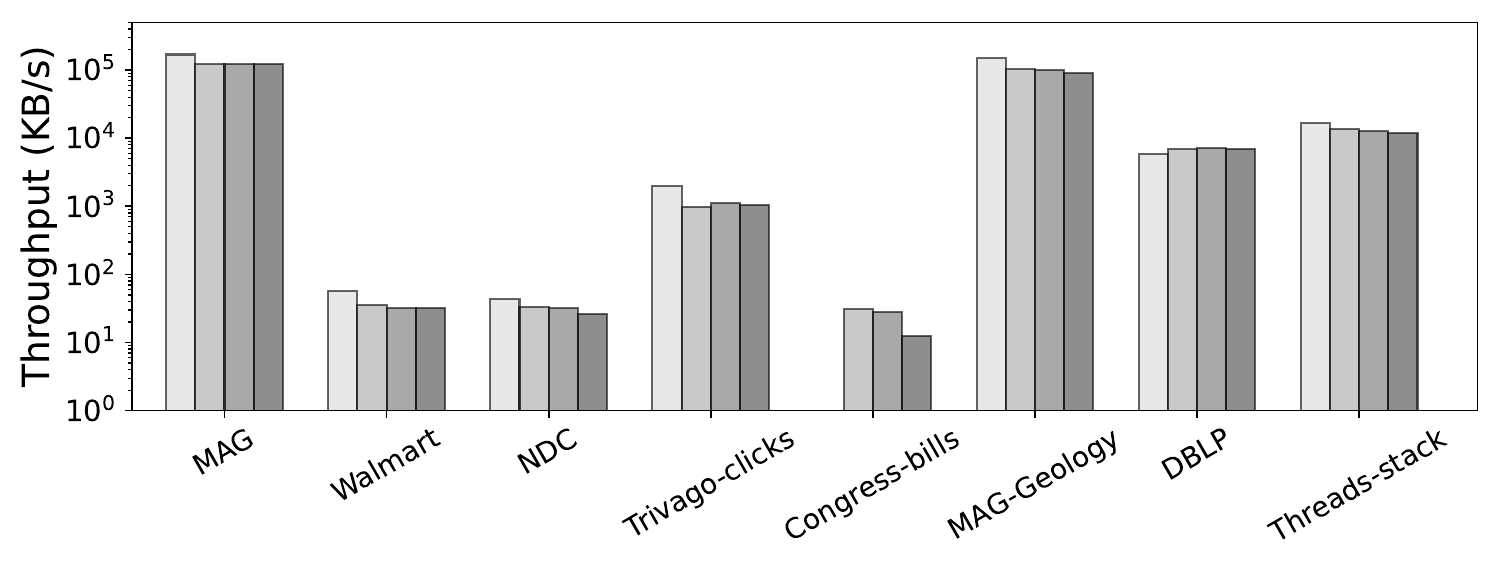}
        \vspace{-0.6em}
    \end{minipage}

    \caption{\revise{Throughput over All Datasets}}
    \label{fig:throughput_dataset}
\end{figure}
\begin{figure*}
    \centering

    \includegraphics[width=0.65\textwidth]
    {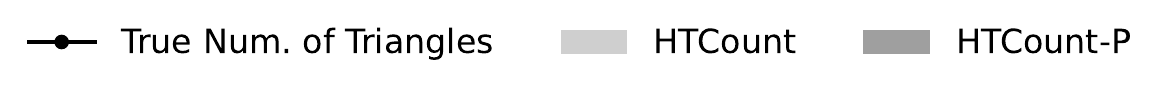}

    
    \begin{minipage}{0.99\textwidth}
        \centering
        \includegraphics[width=0.99\textwidth]{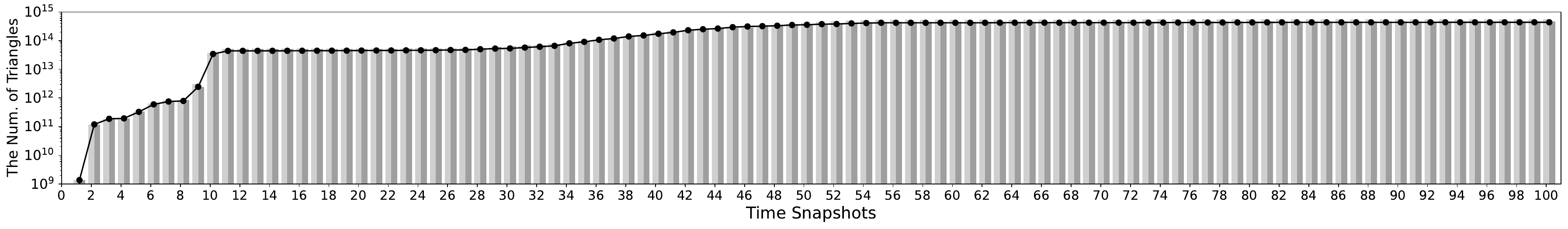}
        \subcaption{\cb}
    \end{minipage}

    
    
    
    
    
    
    
    \begin{minipage}{0.99\textwidth}
        \centering
        \includegraphics[width=0.99\textwidth]{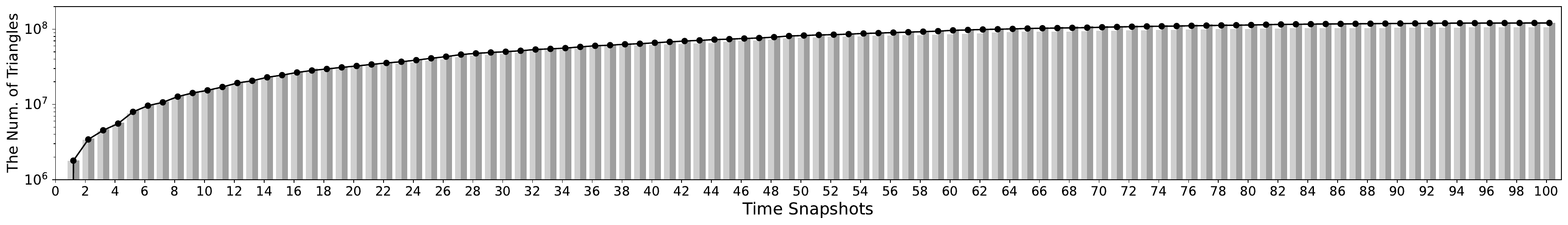}
        \subcaption{\dblp}
    \end{minipage}
    
    


    
    \caption{The Estimated Number of Triangles over Time}
    \label{fig:bc_over_time}
\end{figure*}

\begin{figure}[t]
    \centering

    \includegraphics[width=0.69\textwidth]{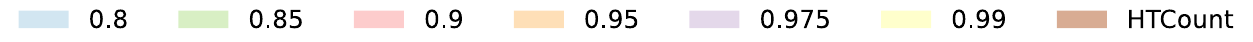}

    
    \begin{minipage}{0.48\textwidth}
        \centering
        \includegraphics[width=\textwidth]{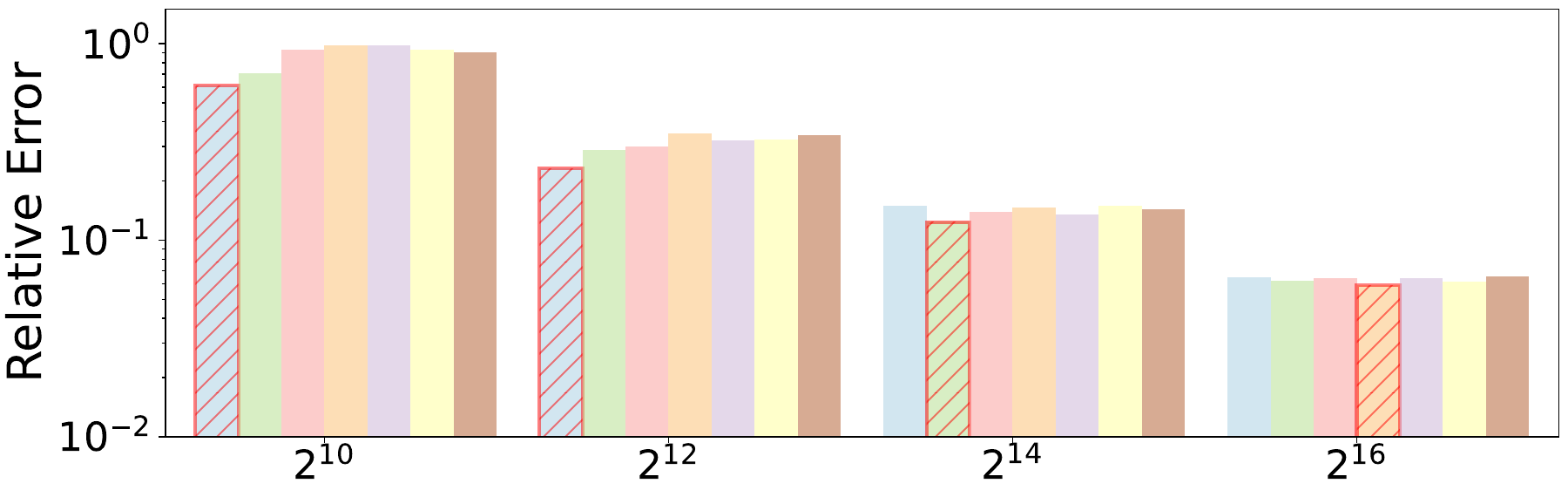}
        \subcaption{\walmart (Sample Size)}
    \end{minipage}
    \begin{minipage}{0.48\textwidth}
        \centering
        \includegraphics[width=\textwidth]{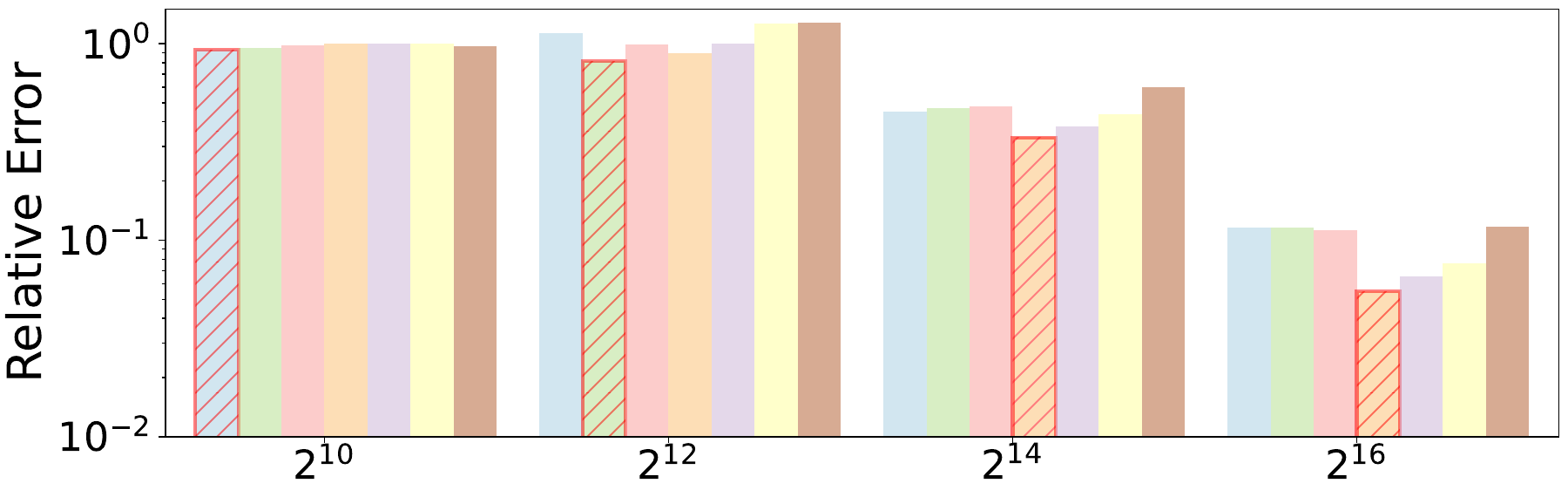}
        \subcaption{\dblp (Sample Size)}
    \end{minipage}
    \caption{\reviseone{Impact of Memory Utilization Threshold $\tau$}}
    \label{fig:alpha}
\end{figure}

\stitle{Exp-1: Accuracy.}
\reviseone{We assess the accuracy of hyper-vertex and hyper-edge triangle estimates across various algorithms and sample sizes ranging from $2^{10}$ to $2^{20}$.} \revisethree{The “sample size” refers to the total number of vertices included in the sampled hyperedges that can be stored, corresponding to a memory budget of 4KB to 4MB (as each vertex is stored as a 32-bit integer). 
We evaluate \hypersv under three settings: optimistic ($\lambda_{|e|_{min}} = \frac{M}{|e|_{min}}$), pessimistic  ($\lambda_{|e|_{max}} = \frac{M}{|e|_{max}}$) and balanced ($\lambda_{|e|_{avg}} = \frac{M}{|e|_{avg}}$ and $\lambda_{2|e|_{avg}} = \frac{M}{2|e|_{avg}}$).
For \alp, the memory utilization threshold parameter $\tau$ is set adaptively according to the sample size. Specifically, for most datasets, we set $\tau=0.85$ for $2^{10}$ and $2^{11}$, $\tau=0.9$ for $2^{12}$ and $2^{13}$, $\tau=0.95$ for $2^{14}$ and $2^{15}$, $\tau=0.975$ for $2^{16}$ and $2^{17}$, and $\tau=0.99$ for $2^{18}$, $2^{19}$, and $2^{20}$. For the \cb dataset, due to its highly skewed hyperedge size distribution, we use lower thresholds: $\tau=0.6$ for $2^{10}$---$2^{12}$, $\tau=0.8$ for $2^{13}$---$2^{15}$, and $\tau=0.9$ for $2^{16}$ and above.
Each experiment is repeated 100 times, and the average relative error is reported. The results are shown in Figure~\ref{fig:exp_accuracy} and Figure~\ref{fig:exp_accuracy2}.}

\revisethree{\al and \alp consistently outperform \hypersv and \hypersvars across all settings. This is because \hypersv uses a fixed sample size, which cannot adapt to varying hyperedge sizes. The optimistic setting ($\lambda_{|e|_{min}}$) fails on all datasets, while the balanced ($\lambda_{|e|_{avg}}$) and adjusted ($\lambda_{2|e|_{avg}}$) settings have limited success, with the latter failing only on \cb. The pessimistic setting ($\lambda_{|e|_{max}}$) always works but wastes memory and performs worse.
While \hypersvars improves over \hypersv, it still has higher relative error than our methods due to its strategy of counting triangles only after the entire sampling process is complete.
\alp outperforms \al because it samples more hyperedges under the same memory constraint, reducing wasted space—especially in datasets with highly variable hyperedge sizes. This is evident in datasets like \cb and \mg.
}



Hyper-edge triangle estimation for CCC, TCC, TTC, and TTT shows similar patterns: \alp consistently has the lowest error. TTT is the most accurately estimated triangle pattern due to its higher frequency (2-3 orders of magnitude more than others). When the sample size is small, it is difficult to capture samples of other triangle patterns. As sample sizes grow, the performance gap narrows, and in some cases, other triangle patterns surpass TTT in accuracy. This trend is consistent with variance analysis results.

\stitle{Exp-2: Memory Utilization.}
\revisethree{We compare memory utilization efficiency among all methods (\hypersv under $\lambda_{|e|_{max}}$ and $\lambda_{2|e|_{avg}}$ settings).
Due to space limitations, we only present results under sample sizes $2^{10}$ and $2^{12}$ (Figure~\ref{fig:wasted_space}), with similar trends for the other settings. }
\revisethree{Our methods significantly outperform \hypersv. The pessimistic setting of \hypersv ($\lambda_{|e|_{max}}$) results in low memory utilization (even below 5\% on some datasets), while the balanced setting ($\lambda_{2|e|_{avg}}$) improves utilization but still keeps it under 50\%. This highlights the inefficiency of fixed-edge sampling, as choosing an optimal sample size is difficult in practice. In contrast, \hypersvars with adaptive reservoir sampling achieves over 90\% utilization, similar to our \al algorithm. Our methods consistently maintain high utilization across all settings, and \alp further improves memory usage by reallocating unused space for additional sampling, especially on datasets with skewed hyperedge sizes. For example, on the \cb dataset, \alp increases utilization by over 77\% at both sample sizes $2^{10}$ and $2^{12}$.}


\revisethree{Figure~\ref{fig:mu_over_time} shows memory utilization over time for the \cb and \dblp datasets with sample size $2^{12}$. To better illustrate the process, we increase the stream rate after memory utilization approaches 100\% (around snapshot 20), allowing more hyperedges per time snapshot. On both datasets, \hypersvars, \al and \alp rapidly reach nearly 100\% utilization, but only \alp remains stable, demonstrating a strong ability to adapt memory allocation for varying hyperedge sizes.
Both \al and \hypersvars maintain high and stable utilization (over 90\%) on \dblp, but experience slight decreases and more fluctuations on \cb after initial saturation.
In contrast, the fixed-sample-size versions of \hypersv (\hypersv-$\lambda_{|e|_{max}}$ and \hypersv-$\lambda_{2|e|_{avg}}$) always show low memory utilization, indicating that the available memory is severely underutilized.
We also track the trend of relative error over time under the same setting. As memory utilization decreases, the overall relative error tends to increase and eventually stabilizes. However, \hypersvars still exhibits a higher relative error than our methods because it counts triangles only after the entire sampling process is complete, with the final results consistent with those shown in Figure~\ref{fig:exp_accuracy}.}

\stitle{Exp-3: Throughput.}
\revisethree{
We compare the throughput of \hypersv (under $\lambda = 2|e|_{avg}$), \hypersvars, \al, and \alp across all datasets (Figure~\ref{fig:throughput_dataset}). Each bar shows the average across multiple runs for all available memory settings. 
Under $\lambda = 2|e|_{avg}$, \hypersv could not run within the memory constraint on the \cb dataset, and thus the result is missing.
Note that \hypersv and \hypersvars perform triangle counting only after the entire sampling process is complete, and thus do not support real-time output. To ensure a fair comparison, we adapt their implementation by updating triangle counts immediately after sampling a hyperedge.
}

\revisethree{
All methods achieve similar throughput overall. \hypersv shows slightly higher throughput in most cases due to sampling fewer hyperedges under the $\lambda_{2|e|_{avg}}$ setting, but this comes at the cost of poor memory utilization and higher estimation error.
\al~and \hypersvars exhibit similar throughput performance, as their sampling strategies result in a comparable number of sampled hyperedges.
\al achieves marginally higher throughput than \alp, which is most pronounced in \cb. This is expected, as \alp introduces additional overhead by maintaining subsets and dynamically assigning incoming hyperedges.}

\reviseone{Dataset density greatly affects throughput. \cb has a larger average hyperedge size (8.7) and far fewer vertices than other datasets, resulting in much more overlap between hyperedges and a substantially greater number of triangles (approximately five orders of magnitude more than in \ma). Since our algorithm prunes pairs of non-intersecting hyperedges (Algorithm~\ref{algo:deabc}, line 26), sparse datasets like \ma allow for more efficient pruning and higher throughput. Conversely, dense overlaps in \cb demand greater computation, reducing throughput. \walmart and \ndc exhibit similar density-throughput relationships.}

\stitle{Exp-4: Tracking Triangle Estimates Over Time.}
We evaluate how well our algorithms track hyper-vertex triangle counts over time using two real-world datasets, \cb and \dblp. Each stream is divided into 100 equal time snapshots, and at each snapshot, we record the estimates from \al and \alp against the ground truth (Figure~\ref{fig:bc_over_time}).

Both algorithms closely follow the true triangle counts over time, demonstrating reliable performance throughout the stream, while \alp consistently achieves higher accuracy.
In \cb (Figure~\ref{fig:bc_over_time}(a)), triangle counts rise sharply in the early stages and stabilize after snapshot 10, while in \dblp (Figure~\ref{fig:bc_over_time}(b)), triangle counts grow steadily. Both algorithms capture these trends well.




\stitle{Exp-5: Impact of $\tau$ on Accuracy.}
\reviseone{As shown in Figure~\ref{fig:alpha}, we examine how varying $\tau$ affects the accuracy of triangle counting in \alp. We test $\tau$ values from 0.8 to 0.99 under different sample sizes, using \al as the baseline. Due to space limitations, results on \walmart and \dblp are shown, with consistent trends observed on other datasets. Configurations with the lowest relative errors are marked in \underline{red} and with \underline{stripes}.}

\alp generally outperforms \al across different 
$\tau$ settings, especially on \dblp. As sample size increases, higher $\tau$ values tend to yield better accuracy. A small $\tau$ prevents the creation of new sample subsets, making \alp behave like \al. Conversely, a large $\tau$ leads to small new sets with high $\frac{m[i]}{|G_s[i]|}$ ratios, which can increase estimation error.
Overall, $\tau$ values between 0.9 and 0.95 offer the best balance of stability and accuracy. For datasets with highly variable hyperedge sizes—where memory is more likely to be wasted—a slightly lower $\tau$ is recommended.





\section{\reviseone{Conclusion and Future Work}}
\label{sec:conclusion}
We study memory-efficient triangle counting in hypergraph streams, introducing a full hyper-vertex triangle classification and two unbiased sampling algorithms, reservoir-based \al and its partition-based variant \alp. These accurately estimate both hyper-vertex and hyper-edge triangles while outperforming prior methods in accuracy and memory efficiency on real datasets.

\reviseone{For future research, there are several directions:$(i)$ \emph{More Higher-Order Motifs.} Extending the proposed methods to count higher-order motifs in hypergraphs, such as four-vertex cliques, to capture more complex interaction patterns. $(ii)$ \emph{Distributed and Parallel Implementations.} Develop distributed/parallel implementations of \al and \alp to process large hypergraph streams in real-time across multiple machines~\cite{jin2025postman,yuan2025gpuscan++,hao2023efficient}. $(iii)$ \emph{Broader Applications.} The comprehensive triangle classification defined in this work can also be applied to other hypergraph algorithms, such as clustering~\cite{meng2022index,xu2007scan,chen2019efficient,gao2023efficient,yuan2025hinscan}, core decomposition~\cite{liu2023distributed} and $k$-truss~\cite{liu2020truss}, to improve their performance and insights by incorporating higher-order interactions.}




\bibliographystyle{ACM-Reference-Format}
\bibliography{main}


\end{document}